\documentclass[12pt,preprint]{aastex}

\shorttitle{Pressure distribution in turbulent thermally bistable flows}
\shortauthors{Gazol, V\'azquez-Semadeni \& Kim}

\begin{document}

\def\gamef{\gamma_{\rm e}}
\def\kfor{k_{\rm for}}
\def\ls{\lambda_{\rm s}}
\def\Leq{\lambda_{\rm eq}}
\def\Peq{P_{\rm eq}}
\def\tc{\tau_{\rm c}}
\def\ts{\tau_{\rm s}}
\def\ttu{\tau_{\rm t}}
\def\VS{V\'azquez-Semadeni}

\title{The pressure distribution in thermally bistable turbulent flows}

\author{Adriana Gazol\altaffilmark{1}, Enrique V\'azquez-Semadeni\altaffilmark{1}}
\affil{Centro de Radioastronom{\'\i}a y Astrof{\'\i}sica, UNAM, A. P. 3-72,
c.p. 58089, Morelia, Michoac\'an, M\'exico }
\email{a.gazol@astrosmo.unam.mx e.vazquez@astrosmo.unam.mx}
\and

\author{Jongsoo Kim\altaffilmark{2}}
\affil{Korea Astronomy and Space Science Institute, 61-1, Hwaam-Dong, Yusong-Ku Taejon 305-348, Korea}
\email{jskim@kasi.re.kr}

\begin{abstract}
We present a systematic numerical study of the effect of
turbulent velocity fluctuations on the thermal pressure distribution in
thermally bistable flows. The turbulent fluctuations are characterized
by their rms Mach number $M$ (with respect to the warm medium) and the
energy injection wavenumber, $\kfor = 1/\ell$, where $\ell$ is the injection size
scale in units of the box size $L=$100 pc. The numerical
simulations employ random turbulent driving generated in Fourier space
rather than star-like heating, in order to allow for precise control of the
parameters. Our range of parameters is $0.5 \leq M \leq 1.25$
and $2 \leq \kfor \leq 16$. 
Our results are consistent with the picture that
as either of these parameters is increased, the local ratio of turbulent
crossing time to cooling time decreases, causing transient structures in
which the effective behavior is intermediate between the thermal-equilibrium
and adiabatic regimes. As a result, the effective polytropic
exponent $\gamef$ of the 
simulations ranges between $\sim  0.2$ to $\sim 1.1$, and the mean pressure of
the diffuse gas is generally reduced below the thermal equilibrium pressure
$\Peq$, while that of the dense gas is increased with respect to
$\Peq$. The fraction of high-density zones ($n > 7.1$ cm$^{-3}$) 
with $P > 10^4$ K cm$^{-3}$ increases from roughly 0.1\% at $\kfor =2$
and $M=0.5$ to roughly 70\% for $\kfor=16$ and $M=1.25$. A preliminary
comparison with 
the recent pressure measurements of Jenkins (2004) in CI favors our case
with $M=0.5$ and $\kfor=2$. In all cases, 
the dynamic range of the pressure in any given density 
interval is larger than one order of magnitude, and the total dynamic
range, summed over the entire density range, typically spans 3--4 orders
of magnitude. The total pressure histogram widens as the Mach number is
increased, and moreover develops near-power-law tails at high (resp.\
low) pressures when $\gamef \lesssim 0.5$ (resp.\ $\gamef \gtrsim
1$), which occurs at $\kfor =2$ (resp.\ $\kfor=16$) in our
simulations. 
The opposite side of the pressure histogram decays rapidly,
in an approximately lognormal form. This behavior resembles that of the
corresponding density histograms, in spite of the large scatter of the
pressure in any given density interval. Our results show that turbulent
advection alone can generate large pressure scatters, with power-law
high-$P$ tails for large-scale driving, and provide validation for
approaches 
attempting to derive the shape of the pressure histogram through a
change of variable from the known form of the density histogram, such as
that performed by Mac Low et al.\ (2004).

\end{abstract}


\section{Introduction}

The atomic interstellar
medium (ISM) is generally believed to be thermally bistable.
This property arises because the neutral
gas is thermally unstable for 300 K $\lesssim T \lesssim$ 5000 K under the
isobaric mode of thermal instability (TI; Field 1965; see also the
review by Meerson 1996), allowing for a
configuration in which gas with temperatures above and below
this range can coexist in  thermal pressure
equilibrium (Field, Goldsmith \& Habing 1969; Wolfire et
al.\ 1995, 2003), being mediated by a thin interface of thickness comparable
to the conductive length (Begelman \& McKee 1990). 
This tendency of the HI gas to naturally segregate in two phases has
long been thought to be the dominant mechanism in forming and
maintaining cold cloudlets of sizes $\sim 0.1$ pc confined by the
thermal pressure of a warm, dilute substrate.

On the other hand, the ISM is known to be globally turbulent, with
numerous kinds of energy sources acting on a wide range of scales (e.g.,
Scalo 1987; Norman \& Ferrara 1996), including spiral arm shocks,
large-scale combined instabilities, supernova and HII-region energy
input, etc. (see, e.g., Mac Low \& Klessen 2004 for a review). The net effect
is to produce and maintain a turbulent velocity dispersion that is
transonic with respect to the warm gas, and supersonic with respect to
the cold medium (e.g., Heiles \& Troland 2003).

In recent years the interplay between TI and 
turbulence has been studied by several groups. Some of the main issues
are: The development and signatures 
of TI in a turbulent ISM (e.g.\ V\'azquez-Semadeni, Gazol \& Scalo 2000;
S\'anchez-Salcedo, V\'azquez-Semadeni \& Gazol
2002; V\'azquez-Semadeni et al.\ 2003), the
triggering of TI by strong compressions in the warm medium (Hennebelle
\& P\'erault 1999, 
2000), the generation of turbulence at the nonlinear stages of
development of TI (Koyama \& Inutsuka 2002; Kritsuk \& Norman 2002), the
production and maintenance of small-scale cold structures (Burkert \& Lin
2000; Koyama \& Inutsuka 
2000, 2002; Audit \& Hennebelle 2005),
the interaction between TI and magneto-rotational instability (Piontek \& 
Ostriker 2004) the production and maintenance of
significant amouts of gas at thermally unstable temperatures (Gazol et
al. 2001; Audit \& 
Hennebelle 2004), as motivated by the reports of
several observational studies (Dickey, Salpeter \& Terzian 1977;
Kalberla, Schwarz \& Goss 1985;  Spitzer \& Fitzpatrick 1995; Fitzpatrick \&
Spitzer 1997; Heiles 2001; Heiles \& Troland 2003; Kanekar et al. 2003),
and the thermal pressure distribution in the turbulent ISM (Audit \&
Hennebelle 2004; de Avillez \& Breitschwerdt 2004; Mac Low et al.\ 2004;
see also \VS\ et al.\ 1995; Korpi et al.\ 1999 for discussions of cases
without thermal bistability.)

Physical discussions of the effect of turbulent velocity fluctuations in
a thermally bistable medium have been given by S\'anchez-Salcedo \&
al. (2002), V\'azquez-Semadeni et al.\ (2003), Wolfire et al.\ (2003),
and Audit \& Hennebelle
(2004). The first two of these works noted that velocity fluctuations
induce perturbations on the gas that can range from behaving adiabatically,
when the turbulent crossing time $\ttu$ across the perturbation size
scale is much shorter than the cooling time $\tc$, to behaving according
to the thermal equilibrium condition between cooling and
heating, in the opposite limit. The turbulent crossing time, in turn,
depends on the scale and amplitude of the velocity fluctuations, and therefore
the higher the Mach number, or the smaller the typical scale of the
turbulence, the higher the fraction 
of fluid parcels that are expected to transiently behave closer to an
adiabatic regime, and 
farther away from thermal equilibrium.  S\'anchez-Salcedo \&
al. (2002) and V\'azquez-Semadeni et al.\ (2003) used this to explain
the presence of significant amounts of gas with temperatures
corresponding to the unstable range. Audit \&
Hennebelle (2004) have further quantified the problem by producing a
semi-analytical model that follows the stretching of a fluid parcel
due to the shearing components of the turbulence while it seeks thermal
equilibrium, to give 
a relation between the amplitude of this component and the amount of
thermally unstable gas present in the flow. Finally, Wolfire et al.\
(2003) have given an estimate of the ratio $\eta$ of the turbulent
crossing time to the cooling time in the warm neutral medium, finding
values 0.3--0.9, which led them to suggest that this medium should often
exhibit non-equilibrium temperatures.

These results have implications for the thermal pressure PDF
in the flow. As a fluid parcel departs from thermal equilibrium, its
pressure also departs from the equilibrium value, and we expect a
\emph{distribution} of the thermal pressure around its thermal
equilibrium value at a given density, determined by the distribution
of Mach numbers of the velocity fluctuations. This holds even in the
absence of direct local heating. 

The  thermal pressure probability density function (PDF) varies
significantly among different models of the ISM. In the simplest equilibrium
multiphase model (Field et al.\ 1969), the thermal pressure PDF is
simply a delta function at the mean value of the
pressure in the midplane. If this value encompasses the unstable range,
then the medium is expected to be segregated into the two phases,
both at the mean, equilibrium pressure,
but with densities and temperatures bracketing the unstable range. The
next level of complexity was added by including supernova heating (Cox
\& Smith 1974; McKee \& Ostriker 1977). In particular, the McKee \&
Ostriker (1977) model implied a piecewise power-law pressure PDF
(Jenkins, Jura \& Lowenstein 1983; see also Mac Low et al.\ 2004) with
slope $-19/9$ for $P\le P_c$ and $-23/9$ for $P>P_c$, where $P_c/k =
10^{3.67}$ cm$^{-3} K$. Moreover, the model predicted no pressures below
the equilibrium pressure of the warm and cold phases. This distribution,
however, follows from a consideration of the
probability of a given point in space belonging to a hot supernova
remnant, and the theoretical evolution of these remnants, not from a
consideration of the local thermodynamic changes in 
the gas due to the turbulent compressions and rarefactions (advection)
that are induced by the supernova energy injection.

Inclusion of advection is naturally accomplished in numerical models of
the star-driven ISM (see \VS\ 2002 for a review). In particular, the
recent papers by Mac Low et al.\ (2004) and de Avillez \&
Breitschwerdt (2004) have discussed the pressure PDF resulting in their
simulations, albeit they appear to obtain different functional forms
for it: Mac Low et al.\ find a lognormal PDF, while de Avillez \&
Breitschwerdt find PDFs that appear closer to a power law, with a slope in fact
not too different from that predicted by McKee \& Ostriker (1977). From
the lognormal shape of their PDF, Mac Low et al.\ conclude that it
originates from the density PDF for an isothermal turbulent flow.
However, in those simulations it is not possible to
disentangle the pressure fluctuations induced purely by turbulent
motions, and those due to direct heating from nearby stellar sources.

Observationally, significant pressure fluctuations have been reported 
in the cold medium. Jenkins et al. 
(1983) found, using {\it Copernicus} observations of CI, variations 
greater than an order of magnitude in the cold gas pressure with small
amounts of gas at up to $P/k=10^5$ K~cm$^{-3}$. More recently, 
Jenkins \& Tripp (2001) used the {\it Space Telescope Imaging Spectrograph} 
({\it STIS}) to confirm this result with better resolved data. Additionally 
they found that their results implied an effective polytropic index for 
the cold gas $\gamma>0.9$, which is larger than the $\gamma=0.72$ derived by
Wolfire et al.\ (1995) for cold gas at thermal equilibrium, mentioning
that this could
be due to the fact that compressed regions may have a cooling time larger
than the dynamical time, so it may behave closer to adiabatically. Finally, 
it is well known that in the local ISM there is an apparent pressure 
imbalance between the hot 
($T\sim10^6$K, $P/k_b\sim 11,000$K~cm$^{-3}$) and the warm 
($T\sim6,700$K, $P/k_b\sim 2,300$K~cm$^{-3}$) gas (see e.g. 
Jenkins 2002; Redfield \& Linsky 2004).

In this paper, we present a systematic study aimed at investigating 
in detail the effects of turbulent velocity fluctuations on a
thermally bistable flow, in particular on the transition from
nearly-adiabatic to near-thermal equilibrium behavior, and on the PDF of
thermal pressure. The simulations we present here are not intended as
accurate models of the ISM, but instead as numerical experiments
allowing us to clearly identify the effects of two fundamental turbulent
parameters, the rms Mach number $M$ and the energy injection scale
(characterized by its wavenumber $\kfor$), on the thermodynamic
response of a thermally bistable flow. For this reason, we have opted
for using Fourier random driving, trading up realism for accurate
control of these parameters, and consider non-magnetic,
non-self-gravitating flows. Also, in order to allow for the large number
of simulations needed to cover a significant range
in parameter space, we have restricted most of the simulations to two
dimensions, although we present a few selected cases in three dimensions
(3D) in order to check that the main trends observed the two-dimensional
runs are preserved in 3D.

The outline of the paper is as follows. In \S \ref{sec:model}
we describe the model used for the simulations and its limitations, and in
\S \ref{sec:test}, we present 
convergence tests for this model. 
In \S \ref{sec:results} we then present the main results, concerning the
effective thermodynamic behavior of the flow and the pressure PDF, both
in global form and in specific density intervals, as the parameters $M$ (\S
\ref{subsec:scale}) and $\kfor$ (\S \ref{subsec:mach}) are varied. 
Then, in \S \ref{sec:disc_implic} we discuss these results in
the context of the simple physical scenario of a transition from
near--thermal-equilibrium to near-adiabaticity (\S \ref{sec:discussion}), 
we show persistence of the main trends in 3D (\S \ref{sec:3d}),
and we discuss some
implications of our results for previous models and simulations
(\S \ref{sec:implications}). Finally, in \S \ref{sec:conc} we give a summary 
and some conclusions.

\section{The Model}\label{sec:model}

\subsection{Prescription} \label{sec:prescrip}
We solve the hydrodynamic equations, including the energy equation, to
simulate a region of $100$ pc on a side, 
with periodic boundary conditions. The simulations discussed in this paper
are in two dimensions (2D) except for those described in \S \ref{sec:3d}.

 The equations are
solved by means of a MUSCL-type scheme (Monotone Upstream-centered Scheme 
for Conservation Laws) with HLL Riemann solvers (Harten, Lax, \& van Leer 
1983; Toro 1999), augmented with model terms for
radiative cooling and background heating, and a prescription for random
turbulent forcing. 
The background heating is taken as a constant 
$\Gamma_0=2.51 \times 10^{-26} {\rm erg}\;{\rm s}^{-1} {\rm H}^{-1}$,
where ``H$^{-1}$'' means ``per Hydrogen atom''. This is the value of the
photo-electric heating rate at density $n = 1$ cm$^{-3}$ reported by
Wolfire et al.\ (1995), and is roughly within half an order of magnitude
of its value throughout the range $10^{-2} {\rm cm}^{-3} \leq n \leq 10^3
{\rm cm}^{-3}$, as reported by those authors. We then use this value to fit
the ``standard'' equilibrium $P$ vs. $\rho$ curve of Wolfire et al.\
(1995) assuming that the background heating is in equilibrium 
with a cooling function that has a piece-wise power-law dependence on the 
temperature. We find (S\'anchez-Salcedo et al. 2002)
\begin{equation}
\Lambda=\left\{ \begin{array}{ll}
                0      &\mbox{$T<15\: K$} \\
          3.42\times 10^{16}T^{2.13} &\mbox{$15\: K\leq T<141\: K$} \\
          9.10\times 10^{18}T        &\mbox{$141\: K\leq T<313\: K$} \\
          1.11\times 10^{20}T^{0.565}&\mbox{$313\: K\leq T<6101\: K$} \\
          2.00\times 10^{8} T^{3.67} &\mbox{$6101\: K\leq T$} 
                        \end{array} \right\}
\label{eq:cooling} 
\end{equation}
where the coefficients have units of erg s$^{-1}$g$^{-2}$cm$^{3}$K$^{-\beta}$,
with $\beta$ being the temperature exponent in the corresponding interval. 
Under thermal equilibrium 
conditions, the gas is thermally unstable under the isobaric mode for 
$313\;{\rm K}< T <6102\;{\rm K}$, and marginally stable 
for $141\;{\rm K}< T <313\;{\rm K}$. In thermal
equilibrium, the transition temperatures  $T=6102$, 313 and 141 K
correspond to densities $n = 0.60$, 3.2 and 7.1 cm$^{-3}$,
respectively. 

The turbulent driving is 100\% solenoidal, and is done in Fourier space
at a specified narrow two- or three-dimensional wavenumber band, 
$k_{\rm for}-1 \leq k  \leq k_{\rm for}$, where $k \equiv
\sqrt{k_x^2 + k_y^2}$ in 2D and $k \equiv \sqrt{k_x^2 + k_y^2 +
k_z^2}$ in 3D, and with the Gaussian deviates having zero mean and
unitary standard  
deviation. The amplitude of velocity perturbations is fixed by a constant 
injection rate of kinetic energy as in the prescription of Mac Low (1999),
although with the difference that we use a different random seed at
each driving time.
The kinetic energy input rate is chosen as to approximately maintain a
desired sonic Mach number.  
 
In the set of simulations presented in \S \ref{sec:results}, the fluid is 
initially at rest and has a uniform density ($n_0=1$ cm$^{-3}$) and 
temperature ($T_0=2399$ K), so that, in the absence of turbulence, the
medium would spontaneously segregate into warm-diffuse ($n=0.34$ cm$^{-3}$, 
$T=7104$ K) and cold-dense ($n=37.2$ cm$^{-3}$, $T=64.5$ K) 
phases. 
The time unit $t_0$ is chosen to be the sound crossing time across the
numerical box at a speed of 9.1 km s$^{-1}$, corresponding to the
isothermal sound
speed at $10^4$ K. Thus, $t_0 = 10.8$ Myr. Mach numbers are expressed
with respect to this sound speed.

For the parameters we use as initial conditions, and in the presence of
realistic thermal conductivity (see \S \ref{sec:test}), 
the maximum linear growth rate occurs at
scales $\sim 8.3$ pc, while the so called Field length (the minimum
unstable scale) is 0.7 pc. In our simulations, thermal conductivity is not 
included, other than the numerical diffusion caused by the
finite-differencing. This causes a ``numerical Field length'' $\sim 3$~pc 
in simulations at a resolution of $512^2$,  
with the maximum growth rate occurring at scales between 12.5 and 25~pc,
as determined through the tests described in \S \ref{sec:test}. 
Thus, the unstable wavelength range in our simulations is somewhat more
compressed than the
real unstable range in the linear regime. We discuss the neglect of
thermal conductivity further in \S \ref{sec:negl_cond}, and resolution
issues in \S \ref{sec:test}.

\subsection{Features and limitations of the model}
\label{sec:limitations}

\subsubsection{Random Fourier driving} \label{sec:Fourier_driving}

In the simulations presented here, the turbulence is driven using a
random scheme executed in Fourier space. This means that every point in
physical space is subject to a force at any given time. This is not a 
very realistic way of driving the turbulence, as in the real ISM the
driving sources, such as supernova explosions or spiral-arm shock
waves are localized in space. Nevertheless, turbulence is expected to
``propagate away'' from the localized sources (Avila-Reese \& \VS\
2001), generating a general turbulent flow. More importantly, we choose
this form of driving because of two reasons. First, it allows us to
precisely control the scale of energy injection as well as the rms Mach
number in the flow. This is
very important, because one of our motivations is to test the scenario
described in S\'anchez-Salcedo et al.\ (2002) and \VS\ et al.\ (2003)
that velocity fluctuations behave closer to an adiabatic regime as the
turbulent crossing time becomes shorter than the cooling time, and the
turbulent crossing time is directly a function of the scale size and
amplitude of the velocity perturbations. 

The second reason is that random Fourier driving
guarantees that all pressure fluctuations that develop are caused purely
by advection (fluid transport) and not by direct injection of heat by
stellar sources, allowing us to isolate the effects of velocity
fluctuations on the pressure distribution. This implies that the
widths of the pressure distributions 
in our simulations should constitute a lower limit
to the actual widths expected in the actual ISM, in which the energy is
injected as heat by stellar sources, directly raising the local
pressure, in addition to any effects of the turbulent advection.

\subsubsection{Neglect of thermal conductivity} \label{sec:negl_cond}

The equations solved in our simulations do not include a model term for
the thermal conductivity. Koyama \& Inutsuka (2004) have suggested that
thermal conductivity should always be included in numerical simulations
of thermally unstable flows, and that enough resolution to resolve the
so called ``Field length'' (Field 1965; Begelman \& McKee 1990) should
be always be used, because of mainly 
two reasons. First, if no thermal conduction at 
all is included, then the TI growth rate asymptotically approaches its
maximum value as the perturbation length scale approaches zero. This
means that, in a finite-resolution numerical grid, the smallest resolved
scale is maximally unstable, regardless of the resolution used, creating
numerical problems. However, in practice this does not occur, as the
numerical diffusion of the code creates a ``numerical Field length'',
that is, a minimum unstable scale larger than the grid cell size, even
if it does not have the same functional temperature dependence as the
actual Field length. Indeed, we have verified that
density perturbations of amplitude 2.5\% and wavelength $\lambda=16$ pixels
(3.1 pc) in a simulation of resolution $512^2$ remain stationary (thus
being the ``numerical Field length''),
while pertubations with $\lambda=4$ pixels (0.8 pc) are completely damped in
times $\sim 3$ Myr. This timescale is
comparable to the $e$-folding time of the growing modes described in \S
\ref{sec:test}. 
Thus, numerical diffusion adequately prevents instability of the
smallest resolved scales. Because of the very small wavelength,
numerical dissipation is more 
effective than TI, which results in damping rather than growing of the
perturbation.

Note that Koyama \& Inutsuka (2004) also
warned that if no realistic thermal conductivity is employed, then the
results are resolution-dependent, because the characteristic scale of
numerical diffusion depends on the resolution. Thus, they concluded
that, in order to have numerically converged results in the presence of
TI, one should i) include an explicit conduction term that is
larger than the conduction due to numerical diffusion, ii) use a cell
size that is smaller than one third of the conductive Field length. 
However, for the \emph{global} properties that interest us here, such
as PDFs of thermal pressure and density, the difference between the
effects of numerical diffusion and of an explicit conduction term, whose
associated Field length is comparable or smaller than the cell size, is
probably not significant. Indeed, in \S \ref{sec:test} we show 
that our simulations are perfectly converged at high densities and
pressures (which is normally the regime of most concern; e.g., Audit \&
Hennebelle 2004) at the resolution we use. 

The second warning of Koyama \& Inutsuka (2004) refers to the
possibility of missing dynamical effects originating from the effects of
the thermal
conduction. Specifically, they describe the generation of motions with
Mach numbers up to 0.13 due to the pressure gradients generated by the
conductivity. Moreover, in their simulations of the development of TI
alone, these motions cause the resulting condensations to coalesce, so
that at the end of the simulation the number of condensations has decreased
by almost a factor of 2. However, Koyama \& Inutsuka (2004) noted that
the initial number of condensations formed in their simulations was
determined by the initial fluctuations, not by the inclusion or omission
of thermal conductivity. In addition, the Mach numbers generated in
their simulations are at least 5 times smaller than those of
the smallest turbulent motions we impose on the flow, and therefore, in
our simulations, density fluctuation production and coalescence is
dominated by the turbulent velocity fluctuations, not by TI in the
presence of thermal conductivity. Thus, for our 
purposes, the motions produced by thermal conductivity can be
neglected.

\section{Convergence test}\label{sec:test}

Here we present results from 20 simulations tailored
to investigate the convergence of the model described in \S
\ref{sec:model}. 

\begin{figure}
\plotone{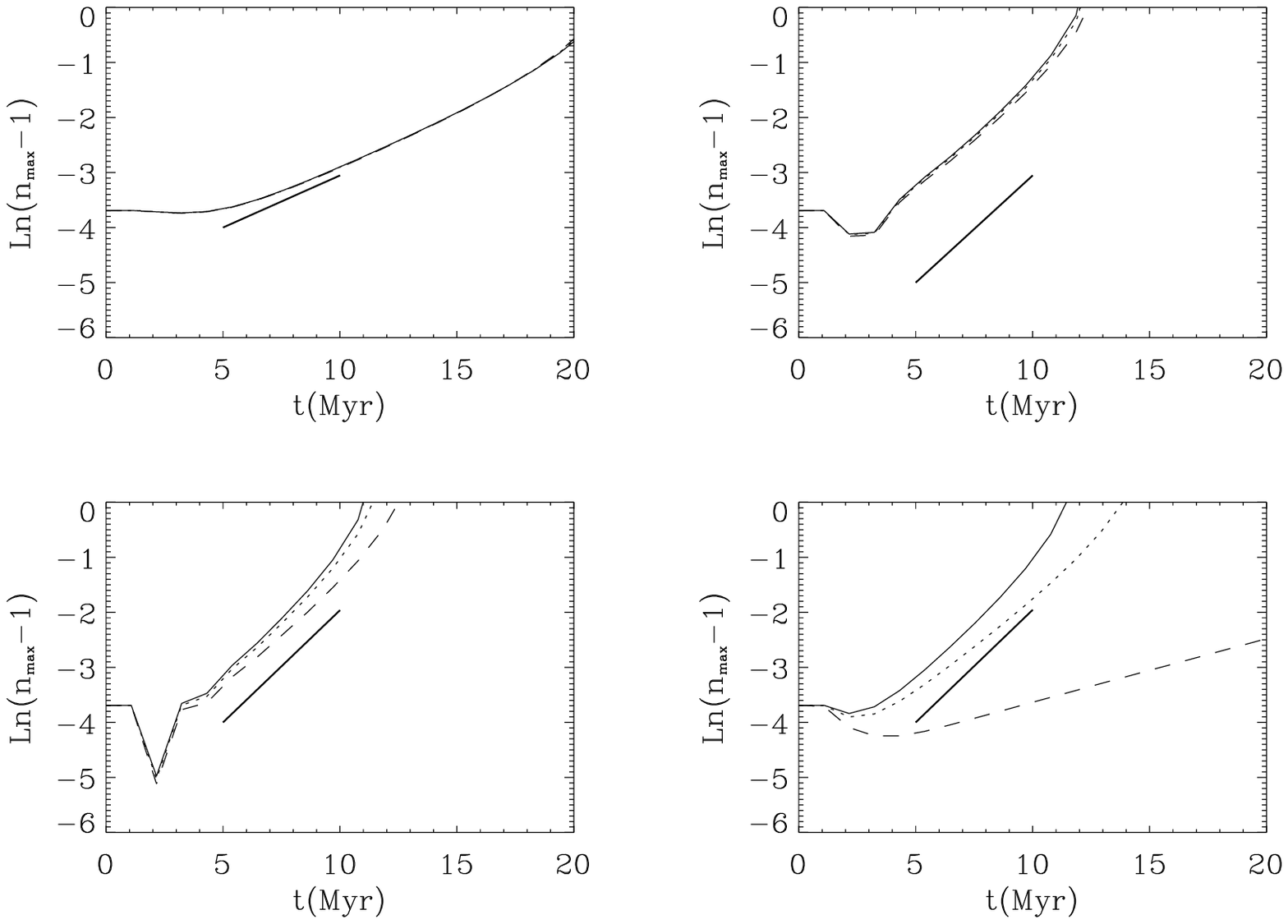}
\caption{Temporal growth of sinusoidal density perturbations with an
initial amplitude of $2.5\%$ and a period of $50$ pc ({\it upper left}),
$25$ pc ({\it upper right})  $12.5$ pc ({\it lower left}) and  $6.25$ pc 
({\it lower right}). The {\it dashed}, {\it dotted} and {\it solid}
lines correspond to resolution of $256^2$, $512^2$ and $1024^2$, 
respectively. The thick
straight lines indicate the slope of the theoretical growth
rate at the corresponding scale.}
\label{fig:tasas1}
\end{figure}

We first consider the linear regime.
In figure \ref{fig:tasas1} we show the temporal growth of sinusoidal
density perturbations with an initial amplitude of $2.5\%$. 
The spatial period of these perturbations is $50$ pc ({\it upper left panel}), 
$25$ pc ({\it upper right panel}), $12.5$ pc ({\it lower left panel}) 
and $6.25$ pc ({\it lower right panel}), 
respectively corresponding to wave-numbers
($k_{\rm p}$) of 2, 4, 8 and 16 respectively. The solid, dotted and dashed
lines are for resolutions of $N=256^2$, $512^2$ and $1024^2$,
respectively. The thick line in each frame indicates the theoretical
growth rate at the corresponding scale. To obtain  
this slope, we have solved the dispersion relation (Field 1965) using
the cooling 
function described above (eq.\ (\ref{eq:cooling})) and a realistic, although 
temperature-independent, value of the conductivity 
$K=K_0 = 5/3(k_b T_0 l/v_{\rm rms})n_0 (3 k_b/2m)$ (Lang 1999), 
where $k_b$ is Boltzmann's constant, and we have taken $T_0=2400$K,
$n_0=1$ cm$^{-3}$, $m=m_H$, $l=3.2\times 10^{-3}$pc and 
$v_{\rm rms} = 5.7$ km s$^{-1}$, the adiabatic sound speed at $T_0$.
These simulations are initially at rest and have a uniform temperature
of $T_0=2400$K. It can be seen that for $k_{\rm p}=2$, 4, and 8 the linear
growth rate of the perturbations in the simulations at all three
resolutions is in good agreement with the theoretical 
growth rate. For $k_{\rm p} = 16$, we see that the run with $N=512$ has
nearly converged to the correct growth rate, while the run with
$N=256$ severely damps the growth of this mode. We conclude that a
resolution $N=512$ is an acceptable resolution for capturing the
linear growth, in the presence of realistic conductivity, of modes with
sizes down to 1/16 the box size.

\begin{figure}
\plotone{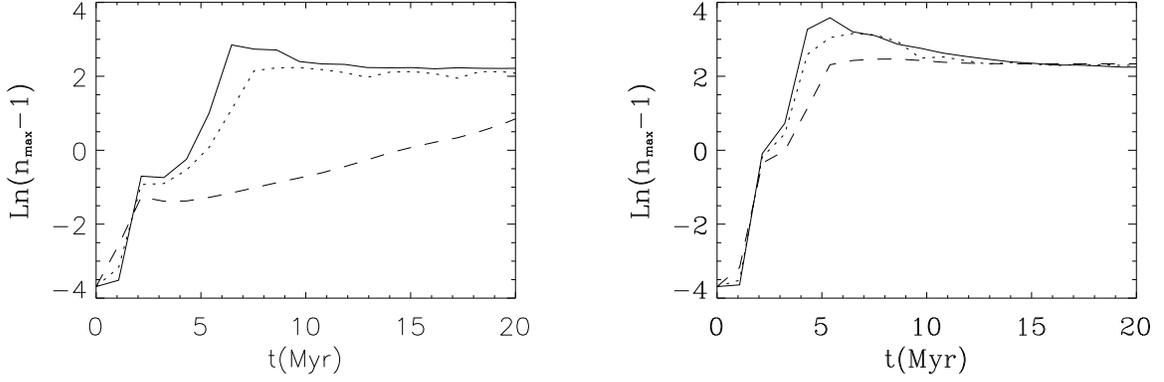}
\caption{Temporal growth of sinusoidal density perturbations with an
initial amplitude of $2.5\%$ and a period of $6.25$ pc ({\it left}) 
and $12.5$ pc ({\it right}) for simulations including an initial 
velocity perturbation at amplitude equivalent to Mach number $M=1.0$ with
respect to the unstable medium at $T=2400$ K. The {\it dashed}, {\it
dotted} and {\it solid} 
lines correspond to resolutions of $256^2$, $512^2$ and $1024^2$, 
respectively.}
\label{fig:tasas2}
\end{figure}

We now turn to the nonlinear case, which is the most relevant one for
the driven-turbulence simulations presented in this paper. 
In figure \ref{fig:tasas2} we display the temporal growth of sinusoidal 
density perturbations with an initial amplitude of $2.5\%$ for simulations 
with sinusoidal large-amplitude (Mach number 1.0 with respect to the
unperturbed medium) velocity perturbations. The 
velocity and density perturbations are in phase, with wavelengths
$6.25$ pc ({\it left}) and $12.5$ pc ({\it right}). In this case the comparison
with the theoretical growth rate is not meaningful. However, it can be seen 
that the difference between the growth rates for $N=512$ and $N=1024$
at $k_{p}=16$ is smaller than for the pure density perturbation case.

\begin{figure}
\plottwo{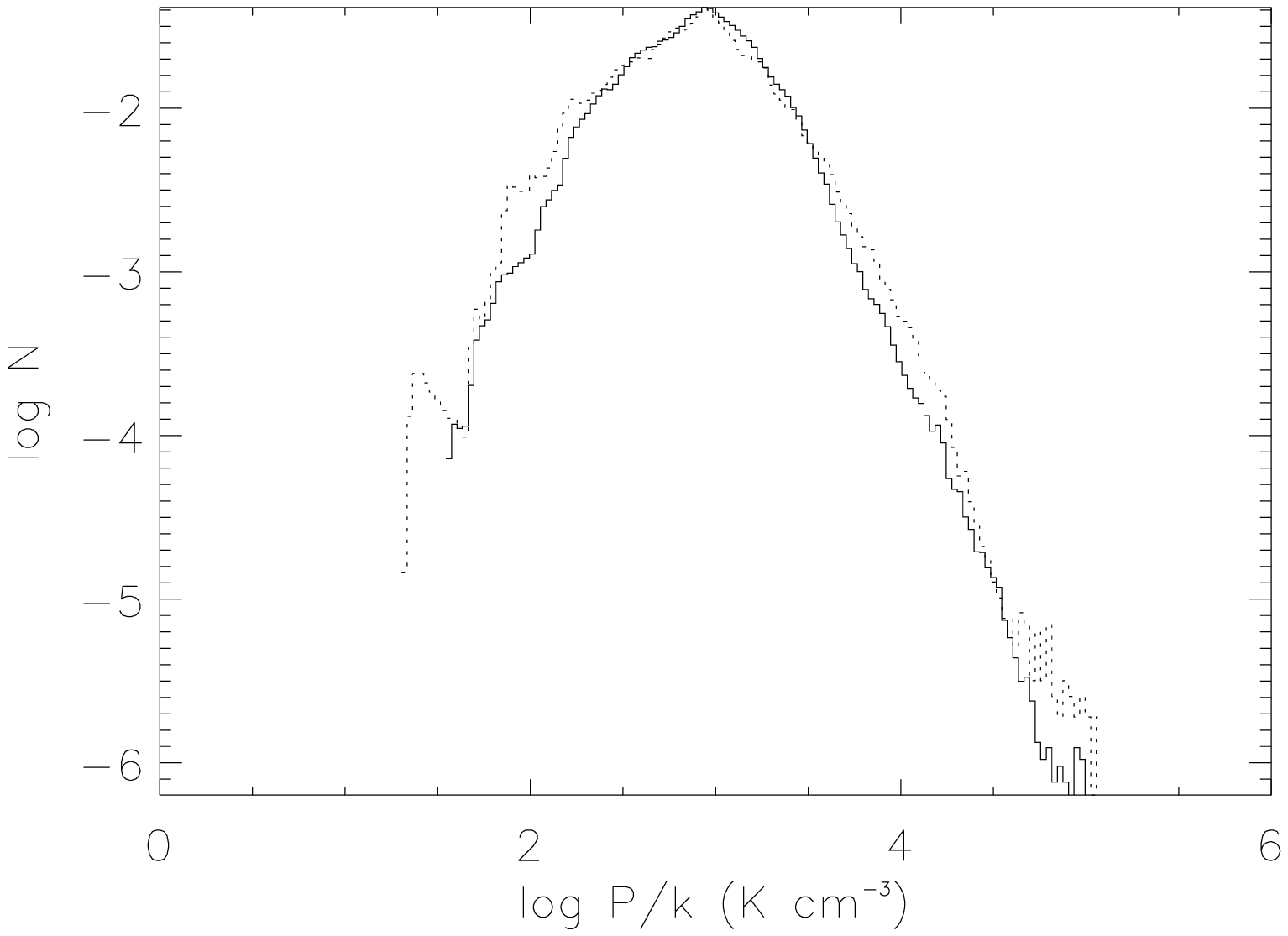}{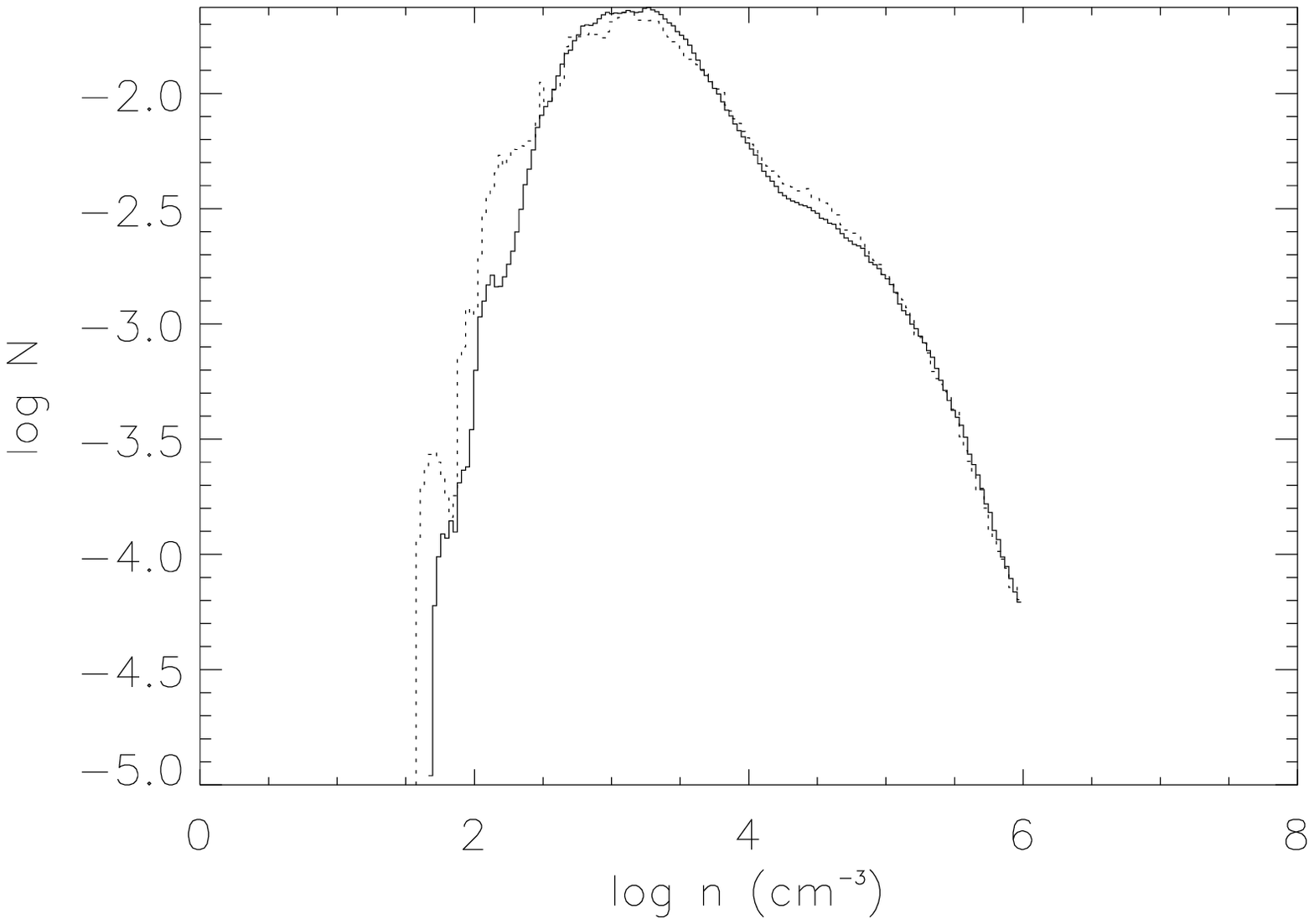}
\caption{Comparison of the total pressure (\emph{left}) and density
(\emph{right}) histograms for simulations with resolutions $512^2$
(\emph{dotted lines}) and $1024^2$ (\emph{solid lines}). The histograms
are normalized to the total number of points, and averaged over the time
interval $1.1 \leq t \leq 2$, where the time unit is the simulation
sound crossing time at $T=10^4$ K}
\label{fig:hist_conv}
\end{figure}

As a final test, we compare the total time-averaged (from 1.1 to 2
simulation sound crossing times) pressure and density
histograms for two fully turbulent simulations with driving
wavenumber $\kfor=2$ and rms Mach number $M=1$, at resolutions of
$512^2$ and $1024^2$ (fig.\ \ref{fig:hist_conv}). We see that the
right sides (high values) of the histograms are perfectly converged at $512^2$,
while the left sides (low values) are approximately so, with the same
relative number of points as the $1024^2$ run in density and pressure
intervals within a factor of 3 from the histogram maximum, and
deviations by factors no larger than $\sim 3$ in more distant
density or pressure intervals. From all of the 
above results, we therefore adopt $N=512$ as a compromise between
acceptable resolution, and the ability to perform the numerous
simulations needed for this study.

Note that with a box of 100 pc on each side and $N=512$, the smallest resolved
scale, even neglecting the effects of numerical diffusion, is $\sim
0.2$ pc. This is larger than the typical size ($\sim 0.1$ pc) of the
cold structures generated by TI and thus our simulations probably
overestimate the sizes of those cloudlets that are formed by
the instability rather than by larger-scale, coherent turbulent
compressions, and certainly do not resolve their internal structure.
However, our interest here is in statistical quantities,
such as the distribution of the pressure values at every density
interval, and the fraction of the volume occupied by gas with a given
pressure. The fact that the high-value sides of the density
and pressure histograms at $512^2$ and $1024^2$ resolutions are nearly
identical suggests that this information deos not require resolving
the tiniest structures, and is accurately captured by our
simulations.

\section{Results}\label{sec:results}
We now describe the results of 10  simulations aimed at characterizing 
the effect of velocity fluctuations on the thermal pressure distribution in 
a thermally bistable flow.  The simulations analyzed
in the next two sections are performed in 2D, with a resolution of
512 grid points per dimension, whereas the simulations described in \S
\ref{sec:3d} are performed in two and three dimensions, with 256 grid
points in each direction. 
The forcing is applied at wavenumbers of $\kfor=2$, 4, 8  
and 16,  implying a driving scale of 50, 25, 12.5 and 6.25 pc respectively,
and has the necessary amplitude to induce turbulent motions with rms
Mach numbers $M\sim 0.5$, 1.0 and 1.25. This is intended to represent
realistic values of the Mach number in the warm phase. Of course,
actual local Mach numbers can be much higher, as the temperature is
generally lower in higher density gas.

Initially, all our simulations are at
rest and have a uniform density ($n_0=1$ cm$^{-3}$) and temperature
($T_0=2400$K). 
In order to allow the simulations to reach a stationary regime, we
evolve the simulations for at least two crossing times across the
turbulence driving scale, given by $t_0/M \kfor$. Specifically, 
cases with $\kfor \ge 4$ are evolved for two code time units ($2 t_0$),
while those with  $\kfor=2$  are evolved for $4t_0$. These times represent
several cooling times at $T_0$, 
given by
\[
\tau_{\rm cool}=\frac{c_{\rm v} T}{\rho_{\rm eq}\Lambda(T)} = 6.27\times10^5
\hbox{ yr},
\]
where $c_{\rm V}$ is the specific heat at constant volume and $\rho_{\rm
eq}$ is the unstable equilibrium density.

\subsection{The effect of the driving scale}\label{subsec:scale}

We first discuss the response of the pressure to changes in the driving
wavenumber $\kfor$, and so in this section we restrict ourselves to rms Mach
numbers $M\approx 1$. Recalling that $M$ is not directly an input
parameter of the simulations, but a result of the energy input rate,
the actual value of $M$ differs slightly from the target
value. Specifically, its actual values are $M=0.97$, 0.96, 0.92 and
0.95 for the runs with $\kfor=$ 2, 4, 8 and 16, respectively.

\begin{figure}
\plotone{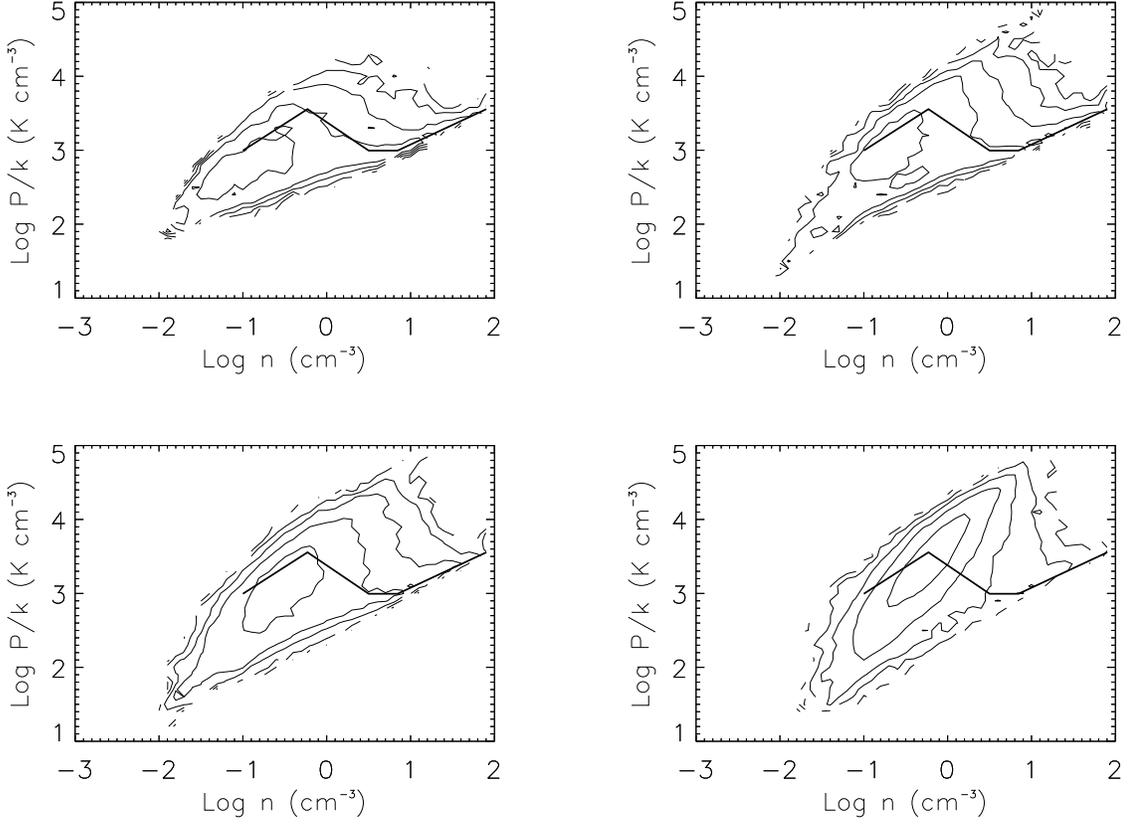}
\caption{Thermal pressure-density relation for simulations with $M\sim 1$ 
and $\kfor=2$ ({\it upper left}), $\kfor=4$ ({\it upper right}), 
$\kfor=8$ ({\it lower left}) and  $\kfor=16$ ({\it lower right}).
The solid line in each panel denotes the thermal-equilibrium pressure.}
\label{fig:pvsrom1}
\end{figure}

Figure \ref{fig:pvsrom1} shows two-dimensional histograms for
these runs, giving the number of grid points in the simulation in a
given $(P,n)$ bin. The contours are logarithmic and are  
set at $10\%$,  $30\%$, $50\%$,  $70\%$ and  $90\%$ of the log$_{10}$ of
the maximum value of the  two-dimensional histogram for each simulation.
The histograms are computed at $t=1.5 t_0$ (resp.\ $t=3.0 t_0$) for
simulations with  $\kfor>2$ (resp.\ $\kfor=2$).
It is clearly seen that as the driving scale 
decreases ($\kfor$ increases), the distribution of points shifts away
from the thermal 
equilibrium curve (denoted by the broken solid line), and towards
adiabatic behavior (with slope 5/3). It is also seen that at low
densities a substantial fraction of the points lies \emph{below} the
thermal equilibrium curve. That is, they have probably been cooled by
negative $P~dV$ work, and have not had time yet to warm back up by the
background heating. Finally, an interesting branch of points seems to
lie on the extension of the equilibrium curve for the dense gas, but at
densities corresponding to the diffuse gas. 

\begin{figure}
\plotone{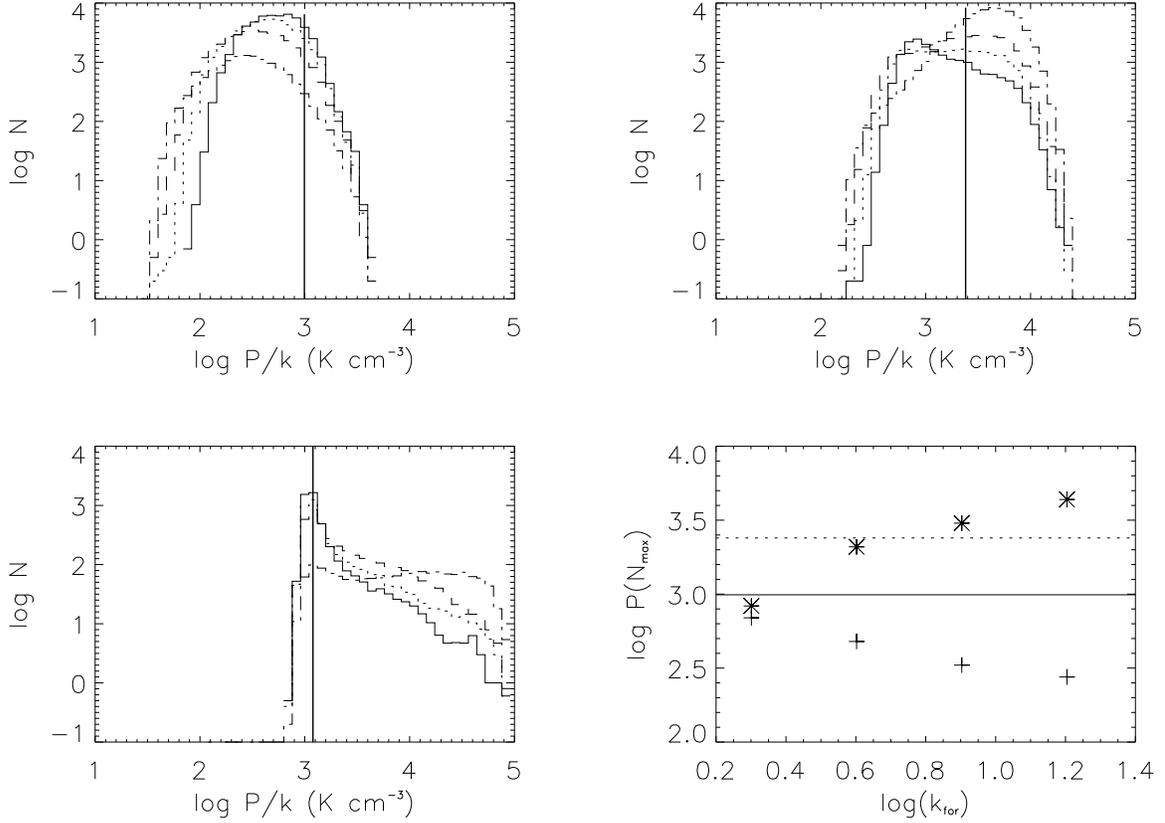}
\caption{Temporally-averaged pressure histograms for simulations with 
$M\sim 1$ and $\kfor=2$ ({\it solid line}), $\kfor=4$ ({\it dotted line}), 
$\kfor=8$ ({\it dashed line}) and  $\kfor=16$ ({\it dashed-dotted
line}). The time averaging is as in fig.\ \ref{fig:hist_conv}.
The histograms are computed over logarithmic density intervals
$n_c/\sqrt{2} \leq n_c \leq \sqrt{2} n_c$,  
centered at $n_c=0.1$ cm$^{-3}$ ({\it upper left panel}),
$n_c=1.0$ cm$^{-3}$ ({\it upper right panel}), and $n_c=10.0$ cm$^{-3}$ 
({\it lower left panel}). The vertical lines denote the
thermal-equilibrium pressure $P_{eq}$ at $n_c$.
{\it Lower right panel}: most probable pressure $P(N_{\rm max})$ for the
histograms centered on   
$n_c=0.1$ cm$^{-3}$ ({\it plus signs}) and 
$n_c=1.0$ cm$^{-3}$ ({\it asterisks}). 
The horizontal lines denote the values of $P_{eq}(n_c=0.1~{\rm
cm}^{-3})$ ({\it 
solid line}) and $P_{eq}(n_c=1~{\rm cm}^{-3})$ ({\it dotted line}).}    
\label{fig:histpromk}
\end{figure}

These trends can also be seen in figure \ref{fig:histpromk}, which shows 
temporally-averaged pressure histograms computed in three density ranges 
$n_c/\sqrt{2} \le n_c < \sqrt{2} n_c$, with 
$n_c=0.1$, 1.0 and 10.0~cm$^{-3}$, corresponding to the warm,
unstable and cold ranges. It can be seen that for low densities the
most probable pressure $P(N_{\rm max})$ is in general \emph{lower} than
the thermal equilibrium pressure at $n_c$ ($P_{\rm eq}$, denoted in
the figure by the vertical lines), and shifts
progressively farther away from it as $\kfor$ increases.
At densities corresponding to the thermally unstable 
range, the most probable pressure increases with increasing $\kfor$,
and passes from being smaller to being
larger than $P_{\rm eq}$. Finally, at $n_c=10.0$~cm$^{-3}$, the four 
histograms peak close to $P_{\rm eq}$, although the height
decreases rapidly with increasing $\kfor$. Moreover, a
high-pressure tail is present, which becomes more populated and more
extended as $\kfor$ is increased.  
Finally, we note that, in all histograms displayed in figure 
\ref{fig:histpromk}, the dynamic range is larger than an order of magnitude. 

\begin{figure}
\plottwo{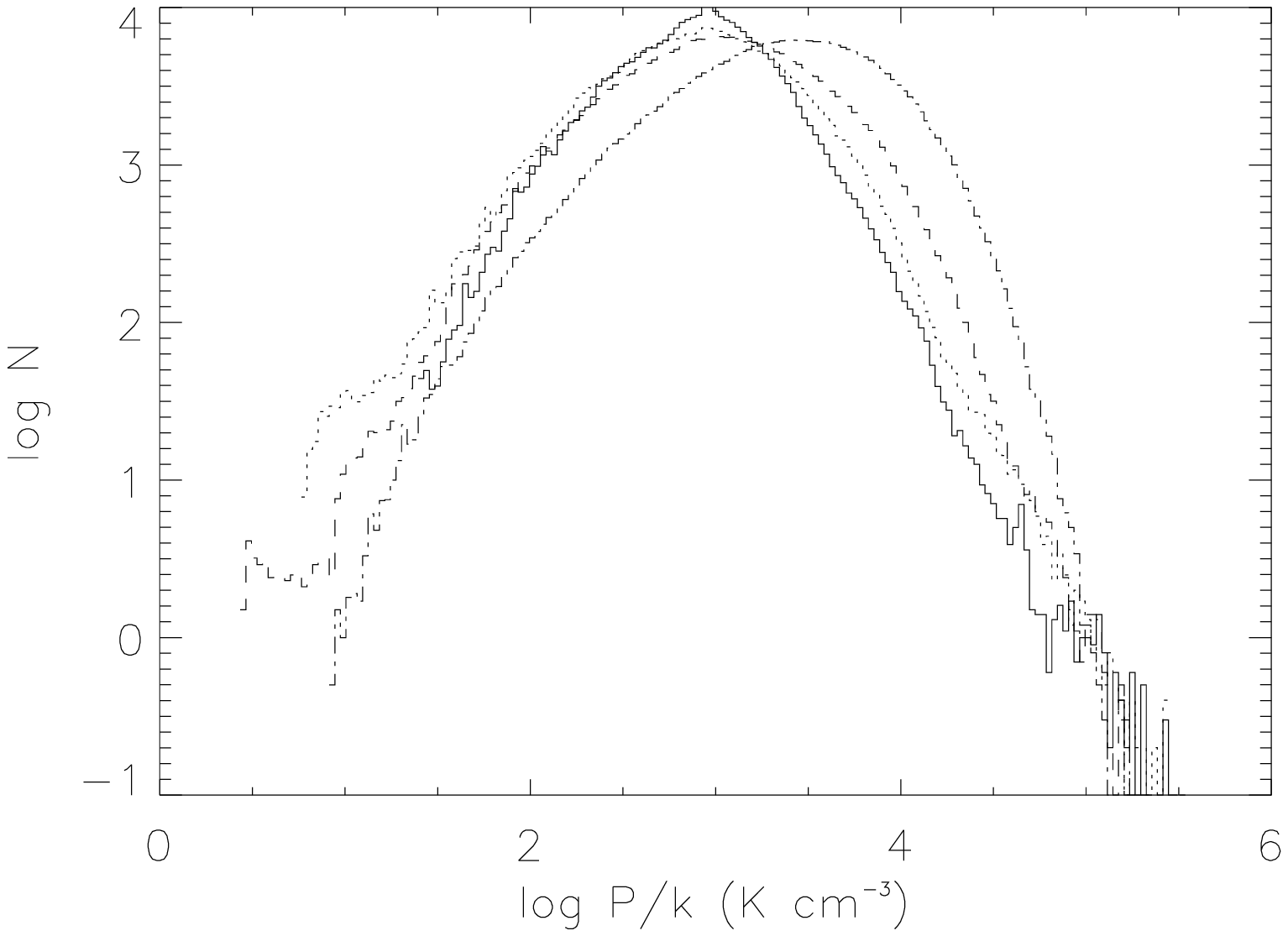}{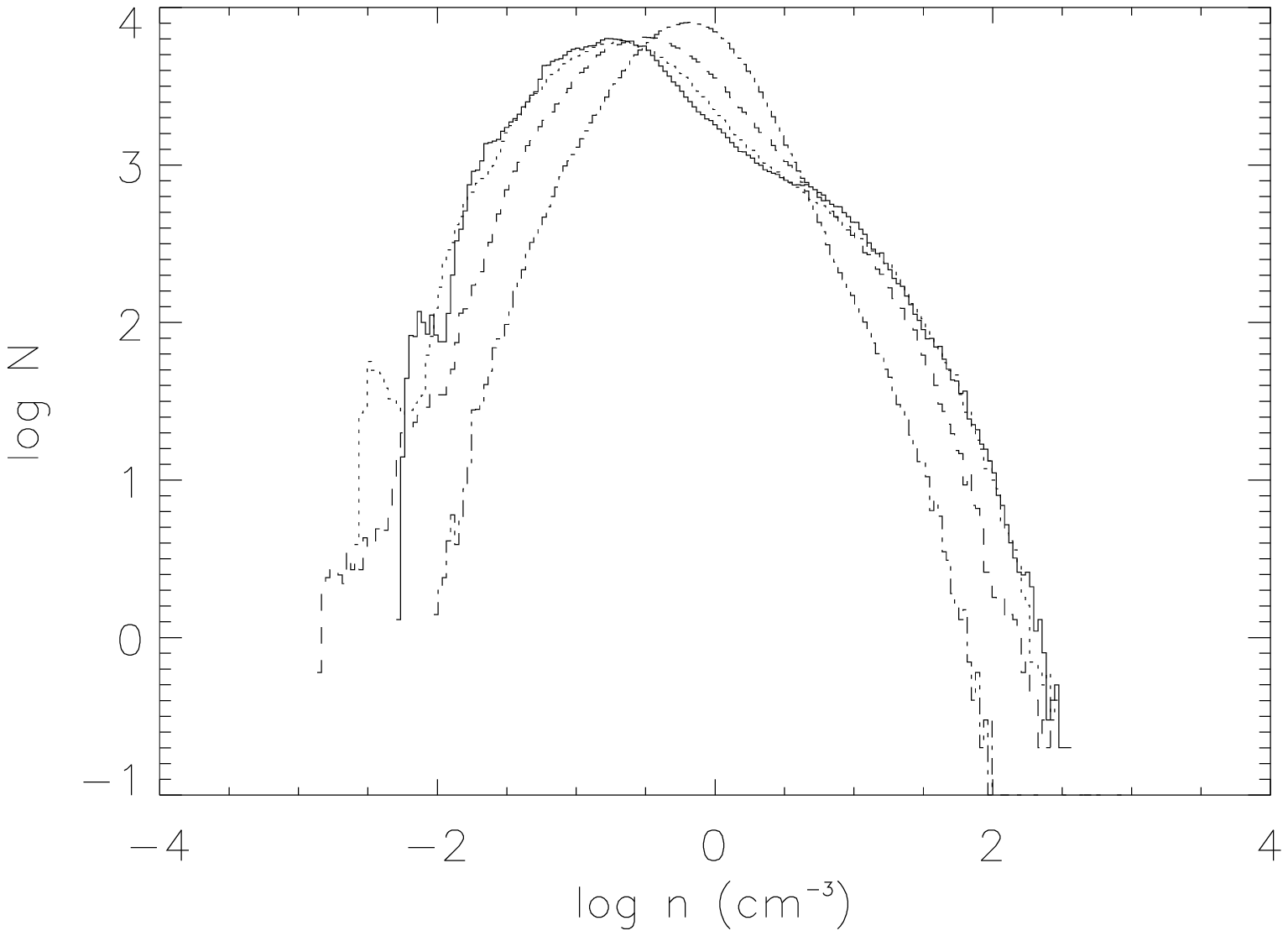}
\caption{Total pressure ({\it left}) and density ({\it right}) 
 histograms for simulations with $M\sim 1$ and $\kfor=2$ ({\it solid line}), 
$\kfor=4$ ({\it dotted line}), $\kfor=8$ ({\it dashed line}) and  $\kfor=16$ 
({\it dashed-dotted line}). The histograms are time-averaged as in fig.\
\ref{fig:hist_conv}.} 
\label{fig:histk}
\end{figure}

The time-averaged pressure and density histograms for the whole field of
all the simulations in this group are shown in figure \ref{fig:histk}. 
The pressure histograms span 3--4 orders of magnitude, with both their
width and $P(N_{\rm max})$ increasing as $\kfor$ increases.
Also, the histograms become more skewed, lifting their low-$P$ side.
The density histograms are
bimodal, but as $\kfor$ increases, the bimodality
becomes less pronounced and the histogram becomes narrower. 

\subsection{The effect of the Mach number}\label{subsec:mach}
We now turn to the effect of the rms Mach number on the pressure
distribution. To this effect, in this 
section we present results from two sets of three simulations each, with  
$M\sim 0.5$, 1.0 and 1.25, and $\kfor=2$ and 16. The actual values
of $M$ in these simulations are $M=0.50$, 0.97  and 1.25  at $\kfor=2$,
and $M=0.56$, 0.95  and 1.25  at $\kfor=16$.

\begin{figure}
\epsscale{.50}
\plotone{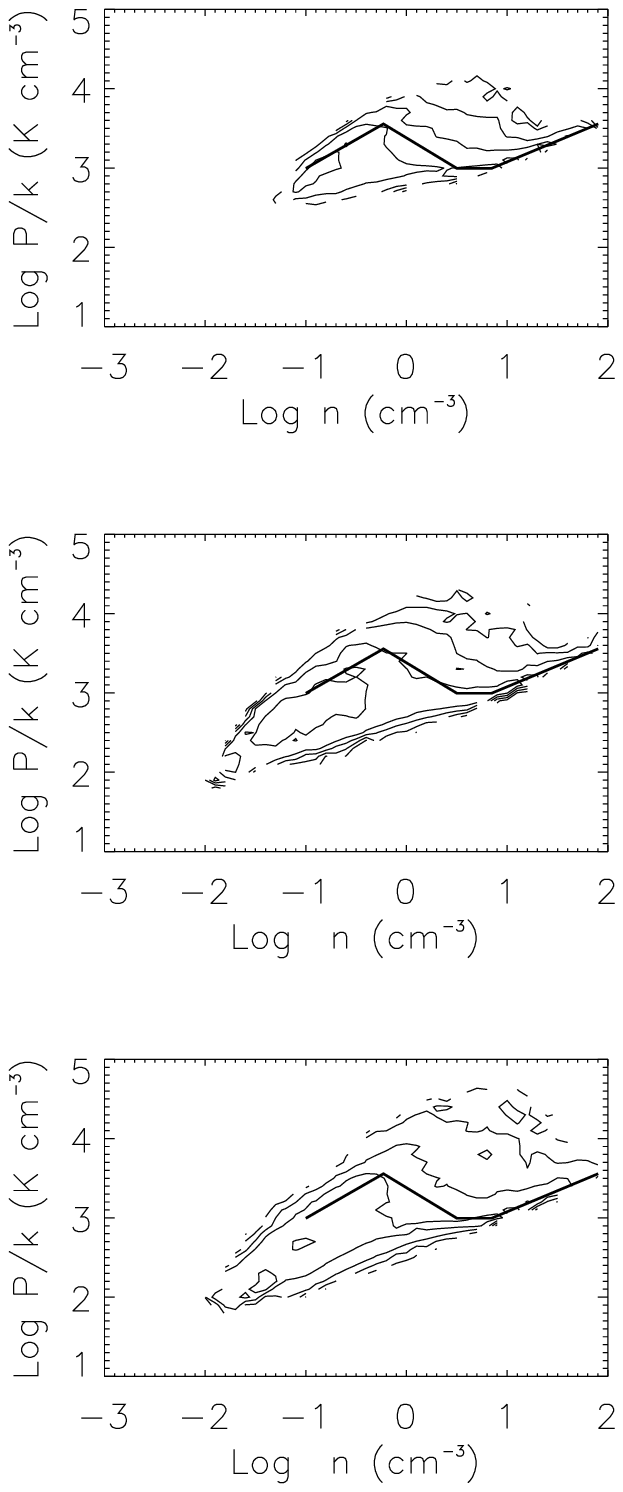}
\caption{Thermal pressure-density relation for simulations with $\kfor=2$ 
and $M\sim 0.5$ ({\it upper panel}), $M\sim 1$ ({\it center panel})
and  $M\sim 1.25$ ({\it lower panel}). The solid line in each panel
shows the thermal-equilibrium pressure.}
\label{fig:pvsrok2}
\end{figure}

\begin{figure}
\epsscale{.50}
\plotone{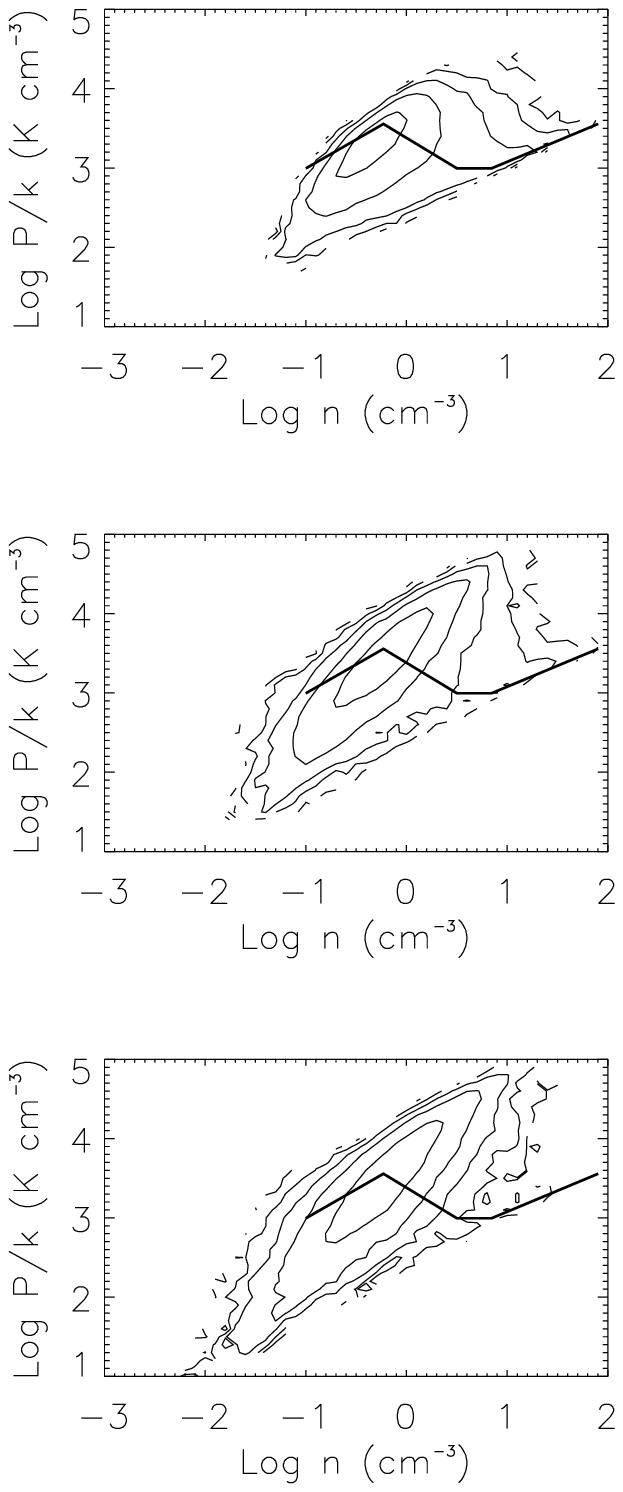}
\caption{Thermal pressure-density relation for simulations with $\kfor=16$ 
and $M\sim 0.5$ ({\it upper panel}), $M\sim 1$ ({\it center panel})
and  $M\sim 1.25$ ({\it lower panel}). The solid line in each panel 
shows the thermal-equilibrium pressure.}
\label{fig:pvsrok16}
\end{figure}

The two-dimensional histograms in the pressure-density space are shown 
in figures \ref{fig:pvsrok2} and  \ref{fig:pvsrok16} for $\kfor=2$ and
16, respectively. The contours are set at the same levels as in \S
\ref{subsec:scale}. The dynamic range of both density and pressure is
seen to increase with increasing $M$ for both driving scales. For
$\kfor=2$, a slight variation of the mean slope of the
distribution of points can be easily seen, while a clear 
steepening of the mean slope is observed at $\kfor=16$. This variation
is summarized in fig.\ \ref{fig:slopes_hiPfrac} for all runs. Finally, the
tendency of the pressure distribution to shift away from 
thermal equilibrium for smaller driving scales, 
reported in the previous section, is seen for all three
values of $M$. 

\begin{figure}
\epsscale{.50}
\plotone{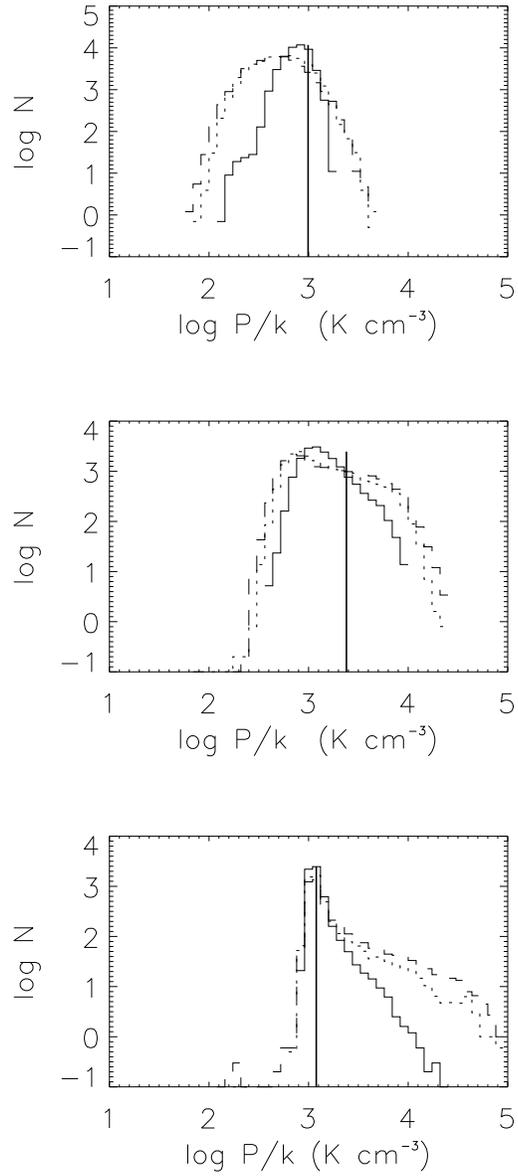}
\caption{Pressure histograms over specific density intervals as in
fig.\ \ref{fig:histpromk}, but for simulations with 
$\kfor=2$ and $M\sim 0.5$ ({\it solid line}),  $M\sim 1.0$ 
({\it dotted line}), and  $M\sim 1.25$ ({\it dashed line}).}
\label{fig:histpromm2}
\end{figure}

The pressure histograms for specific density intervals show a variety of
behaviors. For $\kfor=2$ (fig.\ \ref{fig:histpromm2}), the histogram
centered on $n_c=0.1$ cm$^{-3}$ shifts from being narrow and peaking
very close to $P_{\rm eq}$ at $M=0.5$ to a wider distribution peaking
below $P_{\rm eq}$ at $M=1$ and 1.25. The histograms for the unstable
gas at $n_c=1$ exhibit almost no shift of their peak $P(N_{\rm
max})$, which is located a half order of magnitude below $P_{\rm eq}$, 
as $M$ is increased, although the high-$P$ tail becomes higher, with a
shallower slope, and extends to higher pressures. In the high-$n$
range, the histograms show almost no change in $P(N_{\rm max})$, but the
high-$P$ tail becomes progressively shallower, and extends up to higher
values, as $M$ is increased.

\begin{figure}
\epsscale{.50}
\plotone{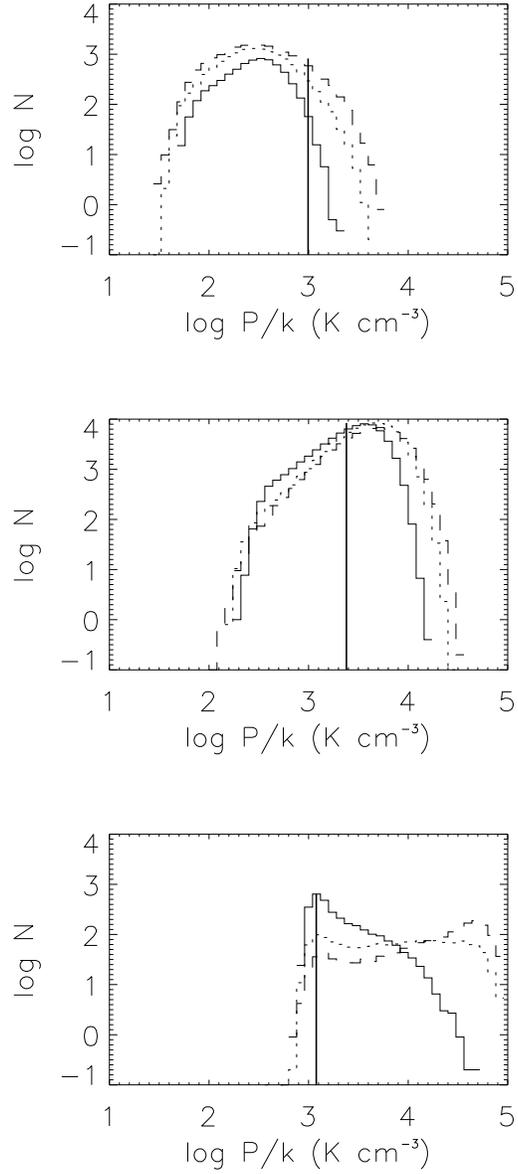}
\caption{Pressure histograms over specific density intervals as in
fig.\ \ref{fig:histpromk}, but for simulations with
$\kfor=16$ and $M\sim 0.5$ ({\it solid line}),  $M\sim 1.0$ 
({\it dotted line}).}
\label{fig:histpromm16}
\end{figure}

For $\kfor=16$ (fig.\ \ref{fig:histpromm16}), the
histograms at $n_c=0.1$ cm$^{-3}$ and 
$n_c=1$ cm$^{-3}$ show little variation in shape and in $P(N_{\rm max})$,
although the histogram at $n_c=0.1$ almost doubles its height, indicating
that a higher fraction of the volume in the simulation is occupied by
diffuse gas as $M$ is increased. This is compensated by a decrease in
the volume occupied by the dense, $n_c=10$ cm$^{-3}$ gas, which
moreover experiences an increment of roughly one and a half orders of
magnitude in its typical pressure $P(N_{\rm max})$.

\begin{figure}
\plottwo{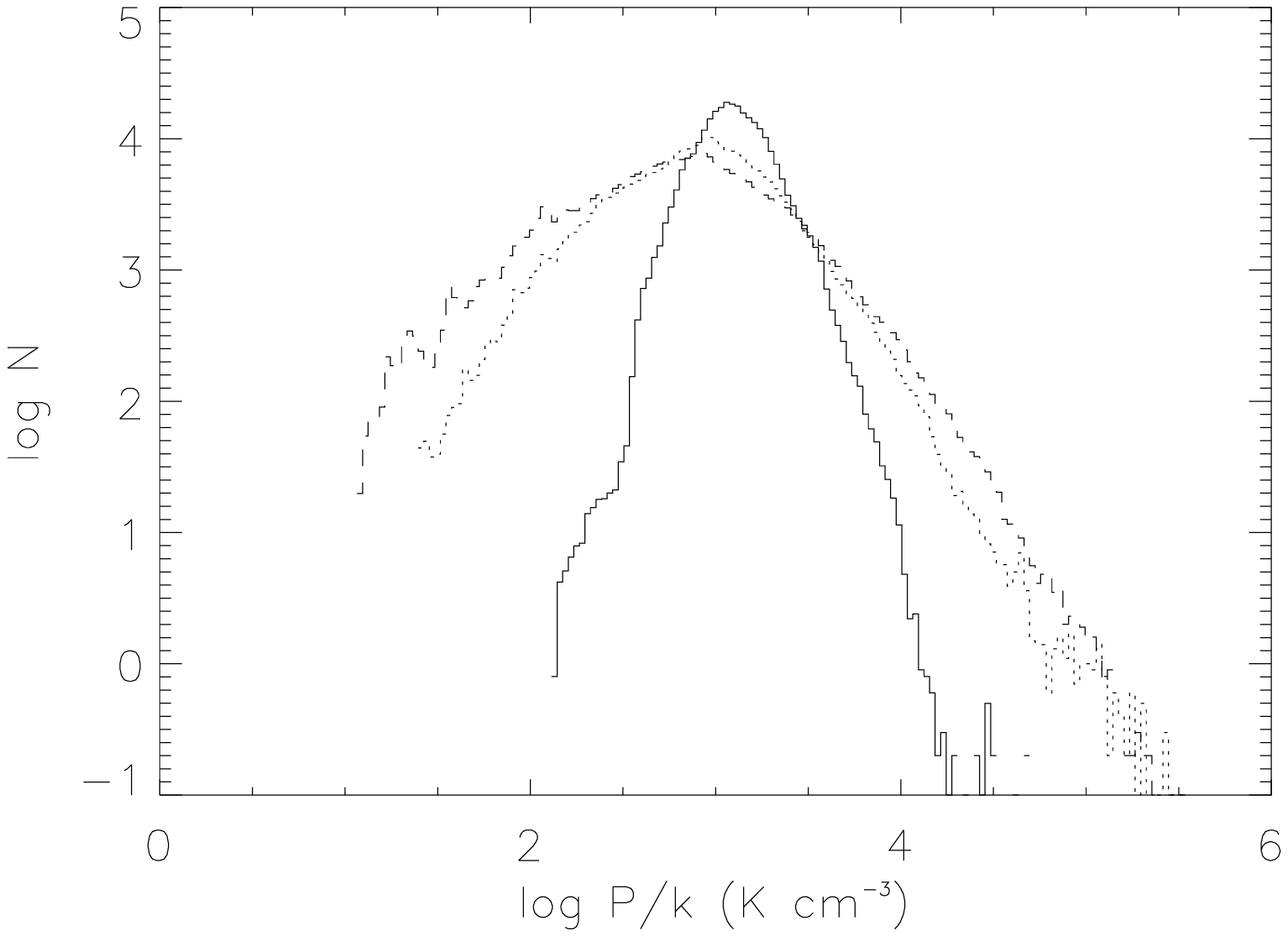}{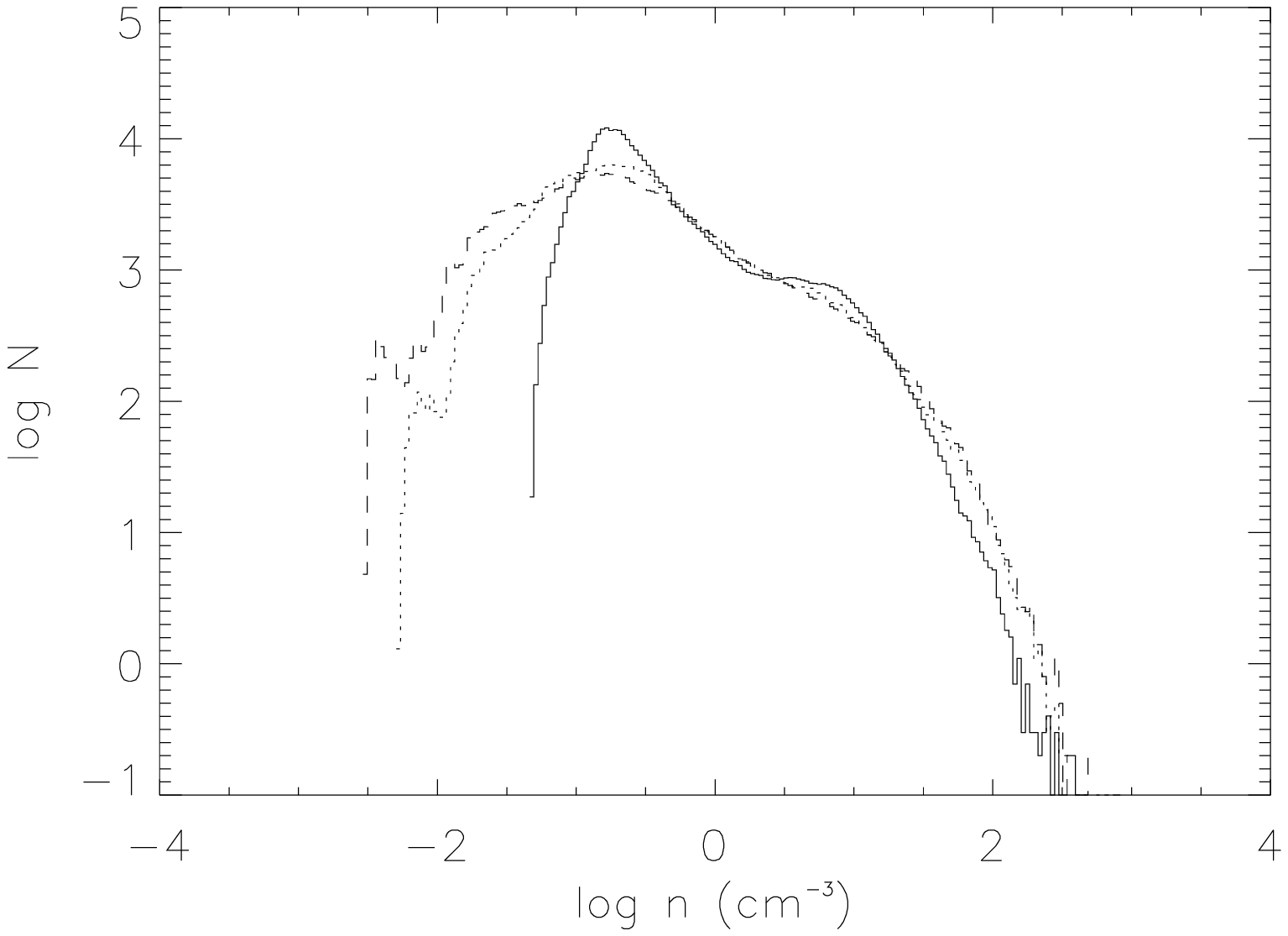}
\caption{Total time-averaged pressure ({\it left}) and density ({\it right})
histograms as in fig. \ref{fig:histk} but for simulations with $\kfor=2$ at
$M\sim 0.5$ ({\it solid line}), $M\sim 1.0$ ({\it dotted line}), 
and  $M\sim 1.25$ ({\it dashed line}).}
\label{fig:histm2}
\end{figure}

In the pressure histograms for the whole simulations
(fig. \ref{fig:histm2} \emph{left}) at $\kfor=2$, the different behavior
between the run with $M \sim 0.5$ and those with
$M \sim 1.0$ and  $M \sim 1.25$ is also evident. In the
former case the histogram is narrower and close to lognormal, although
with a high-$P$ tail for $P \gtrsim 10^4$ K cm$^{-3}$. As $M$ increases,
$P(N_{\rm max})$ shifts to slightly lower pressures, and
the high-$P$ tail reaches closer to the histogram peak, lifting the
entire high-$P$ side of the histogram and causing it to approximate a
power law. On the 
other hand, the low-$P$ side of the histogram widens with increasing
$M$, but never seems to lose its approximately lognormal shape.

In the density histograms for $\kfor=2$ (fig. \ref{fig:histm2}
\emph{right}), the distinction between $M \sim 
0.5$ and the other two cases is also noticeable. The bimodal shape of the
histogram becomes less pronounced as $M$ is increased, with $n(N_{\rm max})$ shifting 
slightly towards lower densities. The histogram width increases
 with increasing $M$, at least in the range explored.

\begin{figure}
\plottwo{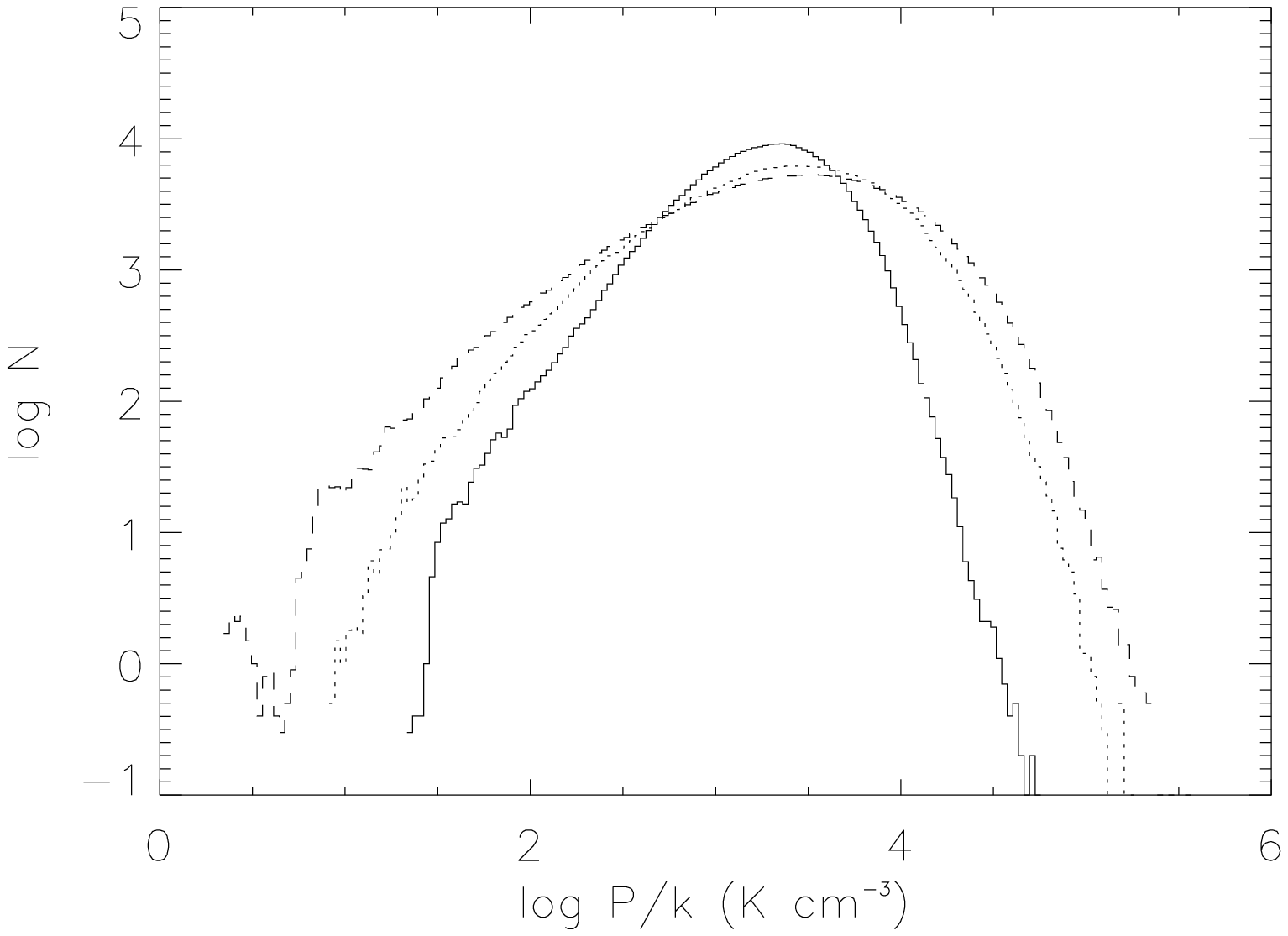}{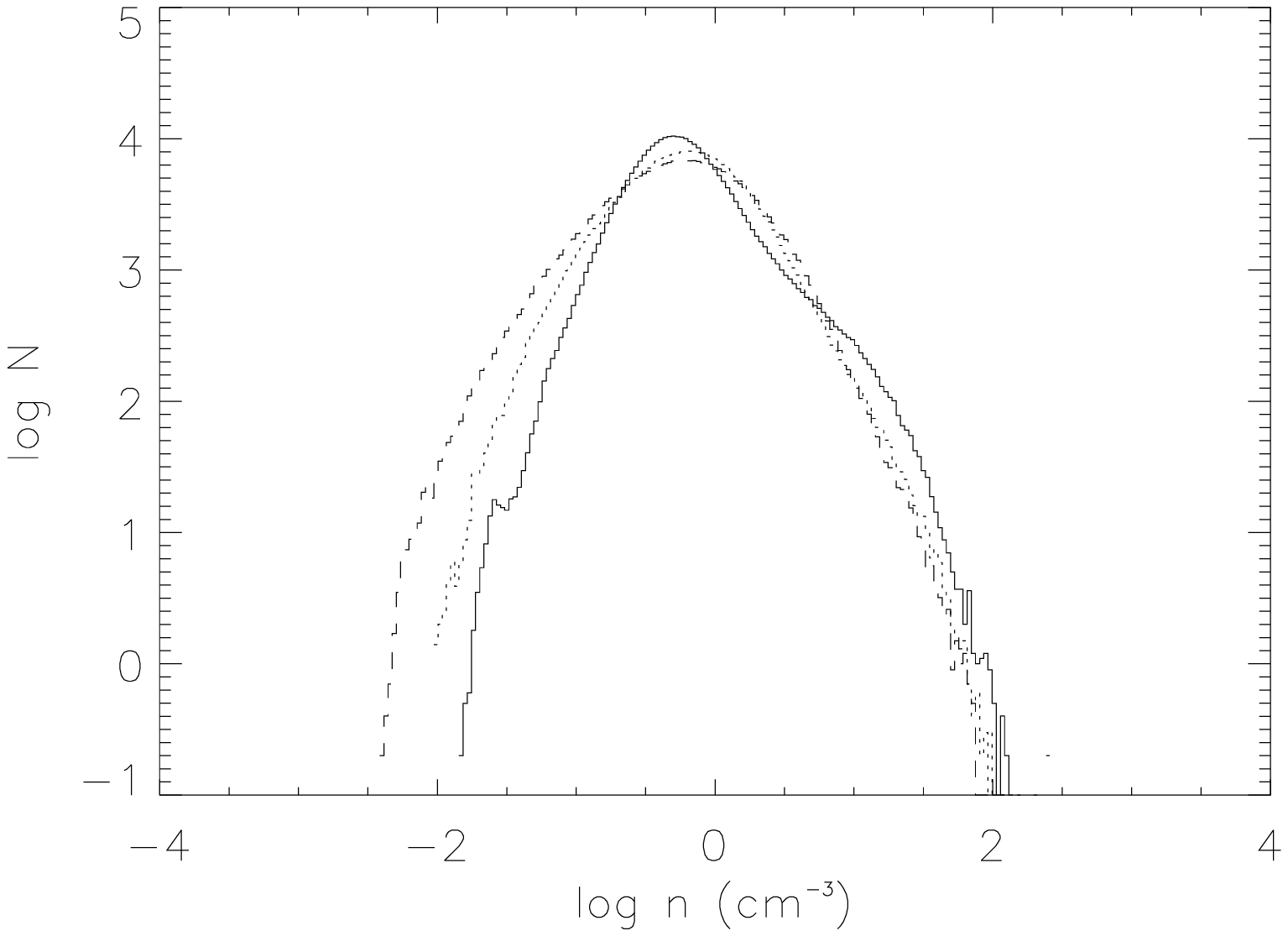}
\caption{Temporally-averaged pressure ({\it left}) and density 
({\it right}) histograms  as in fig. \ref{fig:histk} but for simulations
with $\kfor=16$ and $M\sim 0.5$ ({\it solid line}),  $M\sim 1.0$ ({\it
dotted line}), and  $M\sim 1.25$ ({\it dashed line}).}
\label{fig:histm16}
\end{figure}

For $\kfor=16$ (fig.\ \ref{fig:histm16}, \emph{left}), the high-$P$
branch in the pressure histogram decays 
faster than the low pressure one at all values of $M$, and its slope is
almost independent of $M$, although $P(N_{\rm max})$ shifts to higher
pressures. In this case again the histogram becomes broader with
increasing $M$, but mainly because the low-$P$ tail is lifted and
extends to lower pressures.

Concerning the density histograms, similarly to the case with
$\kfor=2$, for $\kfor=16$, an increase in $M$ leads to  
a less bimodal density histogram  (fig.\ \ref{fig:histm16},
\emph{right}), but in this case, the histogram broadens with increasing
$M$, although mainly by lifting its low-$n$ tail. 
A comparison between the density histograms shown in figures 
\ref{fig:histm2} 
and  \ref{fig:histm16} confirms the fact, discussed in previous section,
that the density distribution becomes narrower with increasing $\kfor$.

\section{Discussion}\label{sec:disc_implic}

\subsection{A unified physical picture} \label{sec:discussion}

The main results of \S \ref{sec:results} can be summarized as follows:
(1) The distribution of points in the $P$-$n$ diagram widens and steepens
as either $M$ or $\kfor$ are increased. 
(2) The mean pressure in a given density interval drifts away from the
equilibrium value $P_{\rm eq}$ as $\kfor$ is
increased, moving towards $P > P_{\rm eq}$ for the dense gas, and towards
$P < P_{\rm eq}$ for the diffuse gas. (3) The pressure histograms in
these density ranges as well as the global pressure histograms are 
generally skewed, and tend to reverse their
skewness with increasing $\kfor$. (4) The pressure histograms in
specific density ranges increase their width as $M$ is increased. (5)
The global pressure histograms develop one near-power-law side and 
one near-lognormal side. For low $\kfor$, the near-power-law develops
on the high-$P$ side, while for high $\kfor$ the near-power-law side
develops at low $P$. In general, the slope of the power law flattens as
$M$ is increased.

Most of these results can be understood simply as a consequence of the
turbulent crossing time becoming shorter in local compressions as
either the Mach number $M$ 
or the driving wavenumber $\kfor$ are increased, creating a larger
fraction of compressions that evolve closer to adiabatically, and
thus temporarily drift away from thermal equilibrium. Indeed, for
our choice of parameters, the cooling and sound crossing times
in the unstable gas are equal at a scale $\Leq \sim 4$ pc. For the warm
medium in thermal equilibrium, this scale increases to $\sim 23$ pc.
The scale $\Leq$ also applies for equality of the 
turbulent crossing time and the cooling time for Mach-1 motions. Below
this scale, classical isobaric perturbations would 
evolve nearly isobarically, because condensation occurs on roughly the
cooling time, which is longer than the time needed to restore pressure
balance (the sound crossing time). However, velocity perturbations
below this scale generate perturbations that approach adiabatic
behavior as their Mach number increases, because in this case the
externally-applied turbulent compression exerts $PdV$ work on the fluid
parcel, heating it on timescales shorter than the cooling
time. Adiabatic perturbations 
are stable to first order (Field 1965), and imply that the pressure
\emph{increases} with increasing density. The net behavior in
transonic flows is expected to be
intermediate between thermal equilibrium and adiabaticity.

\begin{figure}
\plottwo{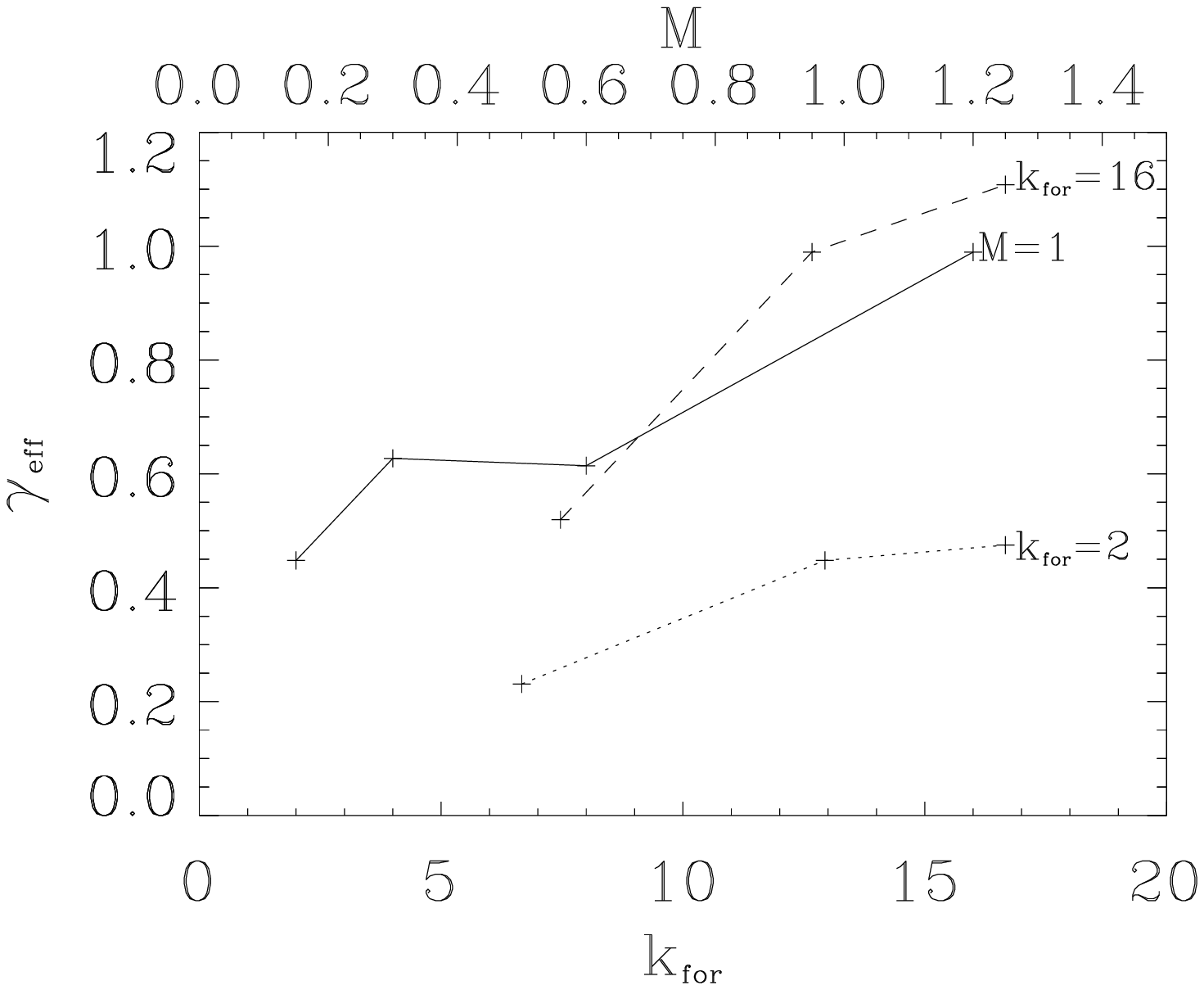}{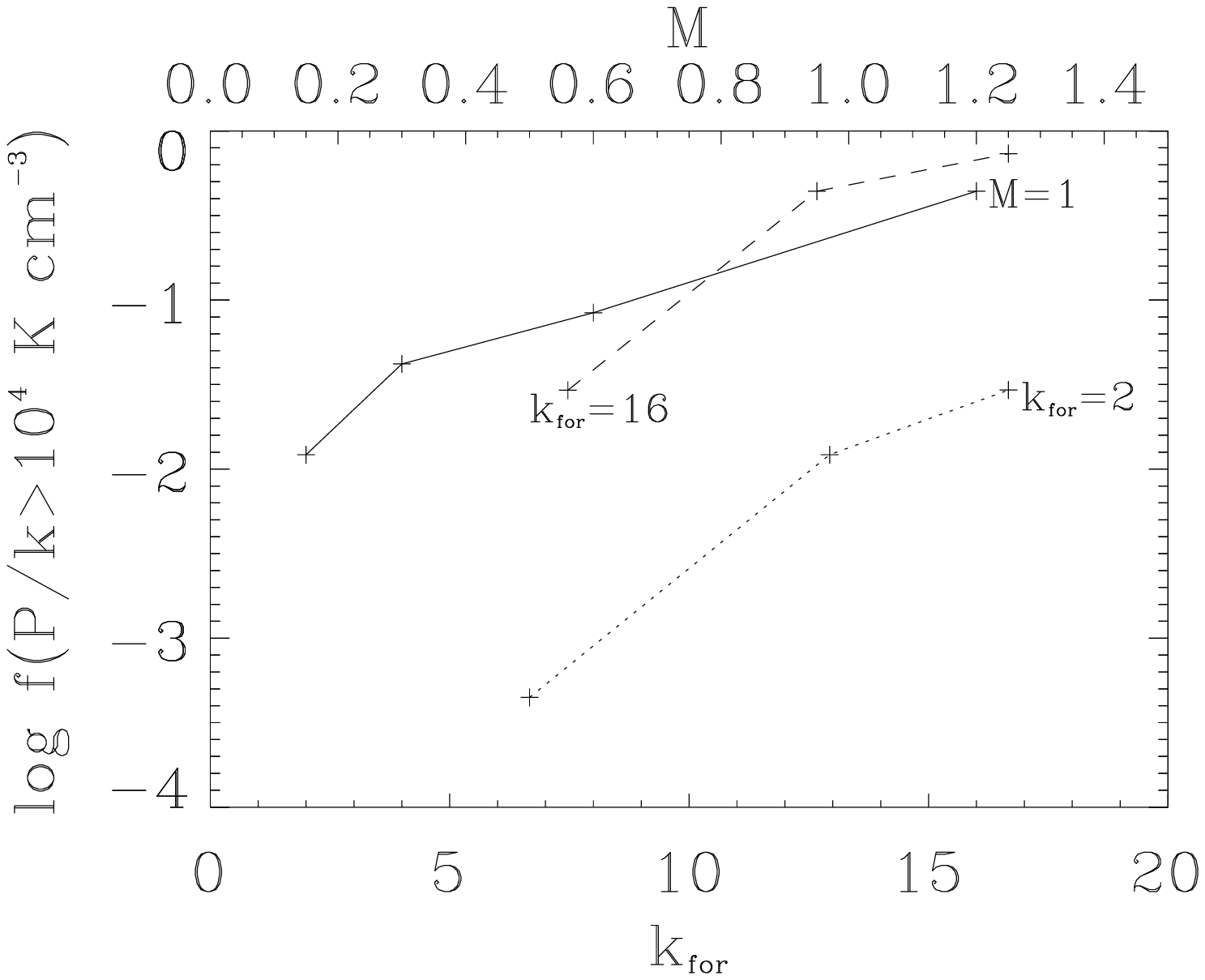}
\caption{\emph{Left:} Least squares slope of the distributions of points in
figs.\ \ref{fig:pvsrom1}, \ref{fig:pvsrok2} and \ref{fig:pvsrok16}. The
{\it solid} line shows the variation with the driving wavenumber
$\kfor$, indicated by the lower horizontal axis, at fixed rms Mach
numbers $M=1$. The {\it dotted} and {\it dashed} lines show the
variation with $M$, indicated by the upper horizontal axis, at a given
$\kfor$, indicated by the label next to  each line.
\emph{Right:} Fraction of grid cells with $P > 10^4$ K cm$^{-3}$  and for the 
$n > 7.1 $~cm$^{-3}$ in figs. The line coding is as in the \emph{left}
panel.}
\label{fig:slopes_hiPfrac}
\end{figure}

The tendency towards adiabatic behavior causes a progressive increase in the
slope (or effective polytropic exponent, $\gamef$) of the ensemble of
points in the $P$-$n$ diagram (fig.\
\ref{fig:slopes_hiPfrac}, \emph{left panel}).  
This causes a \emph{decrease} in the mean pressure of the diffuse
gas, and an \emph{increase} in the mean pressure of the dense gas,
because the point distribution is centered in the unstable range. For
the dense gas, this furthermore causes a tendency towards producing
flatter-topped histograms, because the fraction of high-pressure zones
increases, but the short cooling time at those densities always
produces a significant fraction (most frequently a majority) of points
near $P_{\rm eq}$. Only for the case with the highest $M$ and $\kfor$
does the peak of the pressure histogram for the dense gas shift to the
high-$P$ part of the histogram. Nevertheless, the fraction of
zones in the simulations with $P \geq 10^4$ K
cm$^{-3}$ is seen to increase 
monotonically with either $M$ or $\kfor$ (fig.\
\ref{fig:slopes_hiPfrac}, \emph{right panel}). The fraction of the total
number of cells with $n > 7.1$ cm$^{-3}$ and $P \geq 10^4$ K
cm$^{-3}$ increases from 0.07\% at $\kfor=2$ and $M=0.5$ to 69\% at
$\kfor=16$ and $M=1.25$.

In addition, larger rms Mach numbers are known to cause wider density
PDFs (Padoan et al.\ 1997; Passot \& \VS\ 1998), with the amplitude
of the density fluctuations depending on $\gamef$
(\VS\ et al.\ 1996). As a consequence, wider pressure PDFs are
also expected for stiffer-than-isobaric (i.e., $\gamef > 0$) behavior,
in which the pressure is positively correlated with the density.

For the density PDF, Passot \& \VS\ (1998; see also Scalo et al.\ 1998;
Nordlund \& Padoan 1999)
showed that the density PDF is lognormal for isothermal flows
($\gamef=1$; see also \VS\ 1994), but develops a power-law tail at
high densities for $\gamef < 1$, and at low densities for $\gamef >1$. 
A similar trend is also observed here in response to the
resulting effective polytropic exponent: simulations with $\kfor = 2$
have $\gamef \lesssim 0.5$ (fig.\ \ref{fig:slopes_hiPfrac}, \emph{left}), 
and their resulting density PDFs are skewed to the left, with shallower
high-$n$ tails that extend further from the PDF peak than the
low-$n$ side (fig.\ \ref{fig:histm2} \emph{right}), although with 
signatures of the bimodality associated with the thermal bistability for
the cases with the lowest-$M$. Instead, simulations with $\kfor = 16$,
have density PDFs in which the
low-$n$ side is shallower and generally more extended, although again
with signatures of bimodality on the high-$n$ side at low $M$ (fig.\
\ref{fig:histm16} \emph{right}). Thus, 
the $\kfor=2$ runs generally behave as if having $\gamef <1$, while
$\kfor=16$ runs behave as if having $\gamef >1$.

Interestingly, this dependence of the density PDF on $\gamef$ is also
apparent in the pressure histograms, and in fact it is even more
pronounced, as can be seen in figs.\ \ref{fig:histm2} and
\ref{fig:histm16} (\emph{left panels}). This is somewhat
surprising, given the large scatter of the $P$-$n$ points in any
given density interval. Naively, one would expect that such a large
scatter would preclude any copying of the density-PDF features into the
pressure histogram. Indeed, most features of the distribution do not
survive the change of variable from density to pressure. This is the
case of the scaling of the histogram width or the shift in 
position of the histogram peak with Mach number. Nevertheless, the
development of a near power-law tail depending on the value of $\gamef$
does seem to be preserved, and even amplified, in the pressure histogram.

\subsection{Three-dimensional tests}\label{sec:3d}
The results from \S\S \ref{subsec:scale} and \ref{subsec:mach} have been
obtained in two-dimensional (2D) simulations exclusively, and it is thus
important to determine whether these results are expected to persist in
three dimensions (3D). The distinction between the 2D and 3D cases has
been extensively discussed by \VS\ (1994) and Avila-Reese \& \VS\ (2001).
These authors have argued that the distinction is less pronounced in the
highly compressible case, in which shocks are an important ingredient in
the dynamics. This is because shocks are essentially one-dimensional
structures, independently of the dimensionality of the global flow.

\begin{figure}
\plottwo{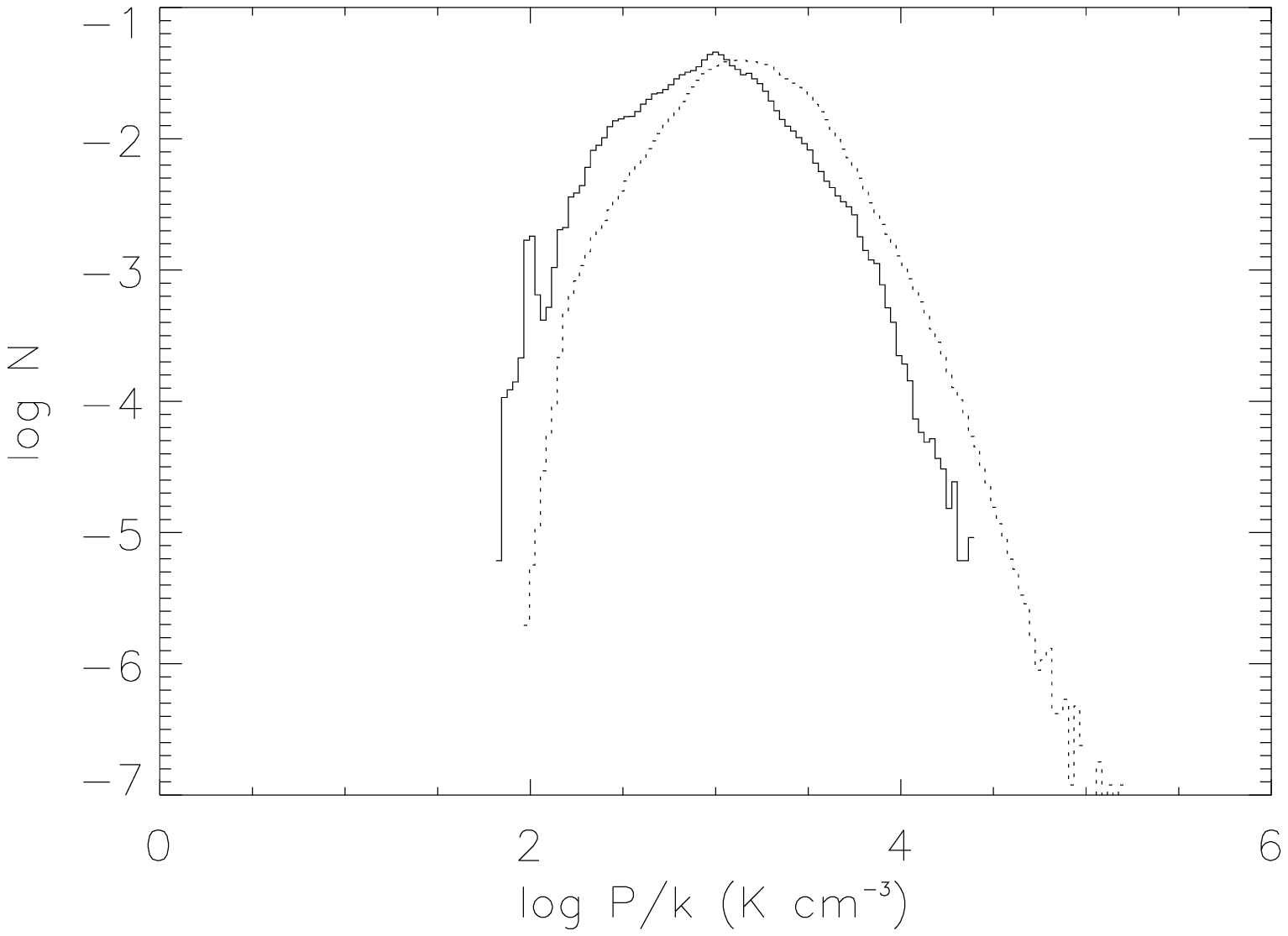}{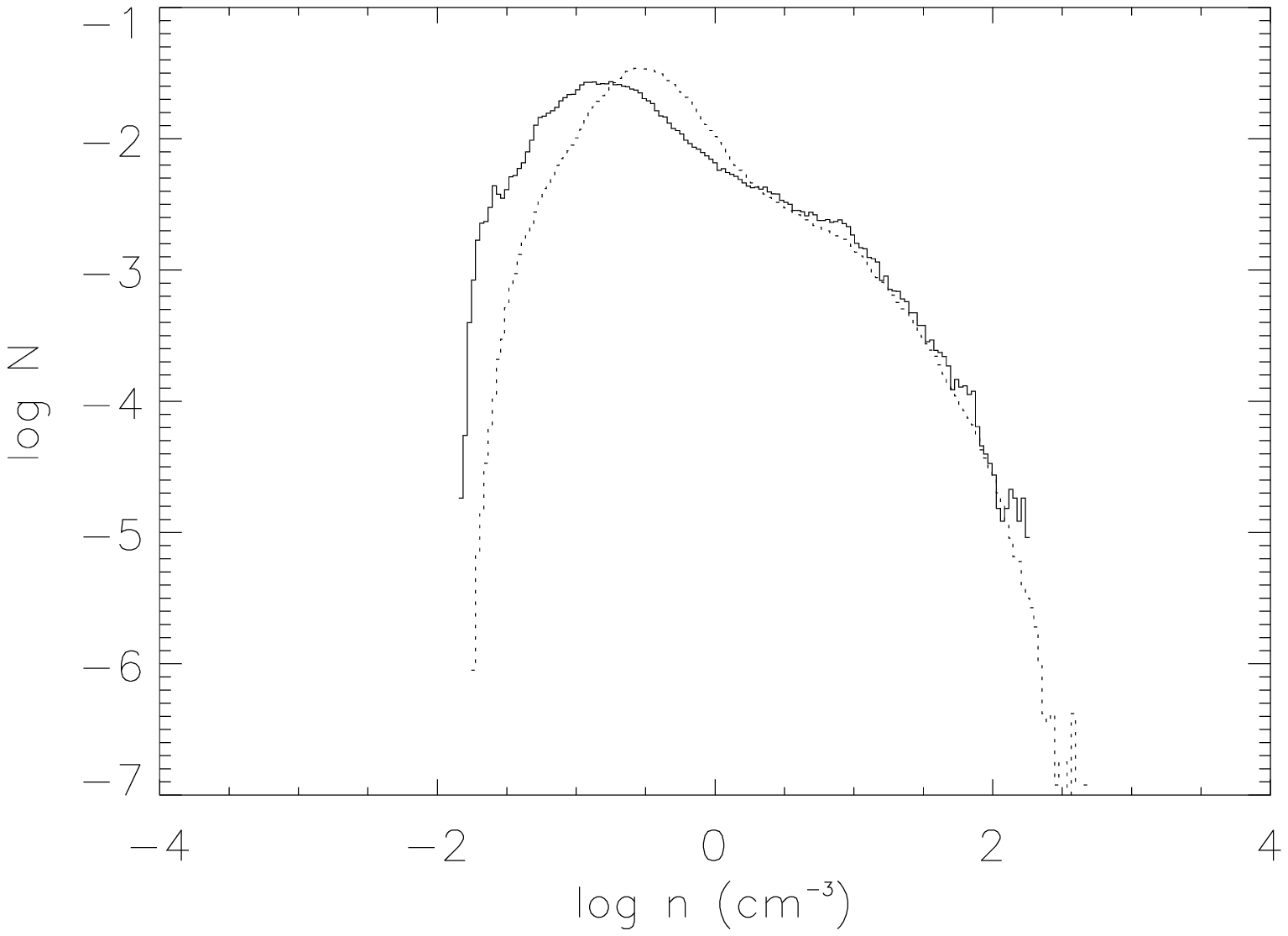}
\caption{Pressure ({\it left}) and density 
({\it right}) histograms  for simulations 
with $\kfor=2$ and $M= 0.90$ in 2D ({\it solid line}),  and 
$M= 0.85$ in 3D ({\it dotted line}). The histograms are normalized to the
total number of points and for the 2D simulations they are
averaged over the time interval $3.2 < t/t_0 < 4$.}
\label{fig:hist3d}
\end{figure}

In order to confirm this expectation in the case of our
particular problem, in
this section we present a comparison between selected cases in 2D and
3D, using simulations with 256 grid points in each direction. The 3D
simulations are performed using a parallel version of the code.  We first
consider two simulations with $\kfor=2$ and $M \sim 0.9$, one in 2D and
the other in 3D. The precise Mach numbers are 0.90 and 0.85,
respectively. Figure \ref{fig:hist3d} shows the total histograms of
pressure ({\it left}) and density ({\it right}).  There we see that the
pressure distribution of the 3D run is systematically shifted to higher
values, by a factor $\lesssim 2$, with respect to that of the 2D one. On the
other hand, the density distributions of the two runs coincide at high
densities, but the 3D case has its peak and its low-density side again
shifted to higher densities by about a factor of 2. 

We speculate that these effects may be due to the fact that the density
peaks are built by compressions of varying dimensionality, up to the
dimensionality of the flow. The higher fraction of low-density
regions in 2D can be understood as a consequence that a peak of a given
average density and size contains a larger fraction of the total mass in
2D than in 3D. Thus, the ``voids'' surrounding the peaks are more
heavily evacuated, giving a higher fraction of low-density zones and
histograms that extend to lower values of the density. This
typically lower density of the voids in 2D can also explain the
typically lower values of the pressure at the low-pressure side of the
distribution. On the other hand, for a density peak formed by a
compression at a given characteristic velocity $\dot R
\equiv dR/dt$, the density varies more rapidly in 3D than in 2D. That
is, assume that $\rho \propto {\cal M} R^{-m}$, where ${\cal M}$ and $R$ are
respectively the mass and radius of the compressed parcel, and $m=2$ in
2D and 3 in 3D. Then $\dot \rho = - m{\cal M} R^{-m-1} \dot R$, and the
density rate of change is larger in 3D than in 2D at a given compression
velocity $\dot R$. Thus, in 3D the characteristic time for variation of
the density is comparatively shorter than the cooling time in 3D, and
the behavior should be slightly closer to adiabatic, explaining the
higher transient pressures in 3D at high densities.

\begin{figure}
\plotone{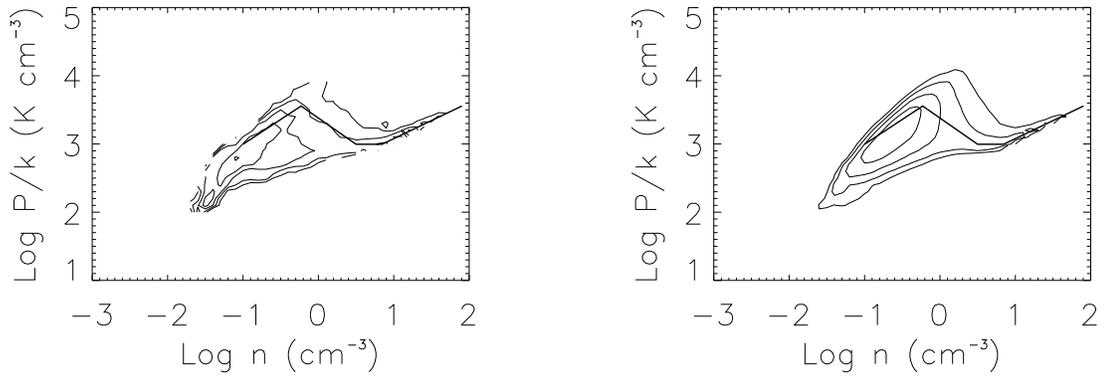}
\caption{Thermal pressure-density relation for three-dimensional simulations with $\kfor=2$ 
 and $M= 0.90$ in 2D ({\it left panel}) and  $M= 0.85$ in 3D 
({\it right panel}). The solid line in each panel
shows the thermal-equilibrium pressure.}
\label{fig:pvsrho3d}
\end{figure}

Nevertheless, the differences in the histograms in 2D and 3D are
relatively minor, probably because the occurrence of high-dimensional
compressions must be a relatively rare event in comparison with
one-dimensional ones (shocks). The pressure distribution in 3D is
shifted by factors not larger than 2 in the 3D case, and in fact the
$P$-$n$ relation is very similar in the 2D and 3D cases, as seen from
two-dimensional histograms in the pressure-density space (fig.\
\ref{fig:pvsrho3d}), with the main difference being that the
distribution of points in the 3D case
extends slightly farther above the thermal equilibrium curve at
densities $n \sim 1$ cm$^{-3}$. The least squares slopes of the
distributions are also very similar, at 0.35 and 0.38 for 2D and 3D,
respectively.
 
\begin{figure}
\plottwo{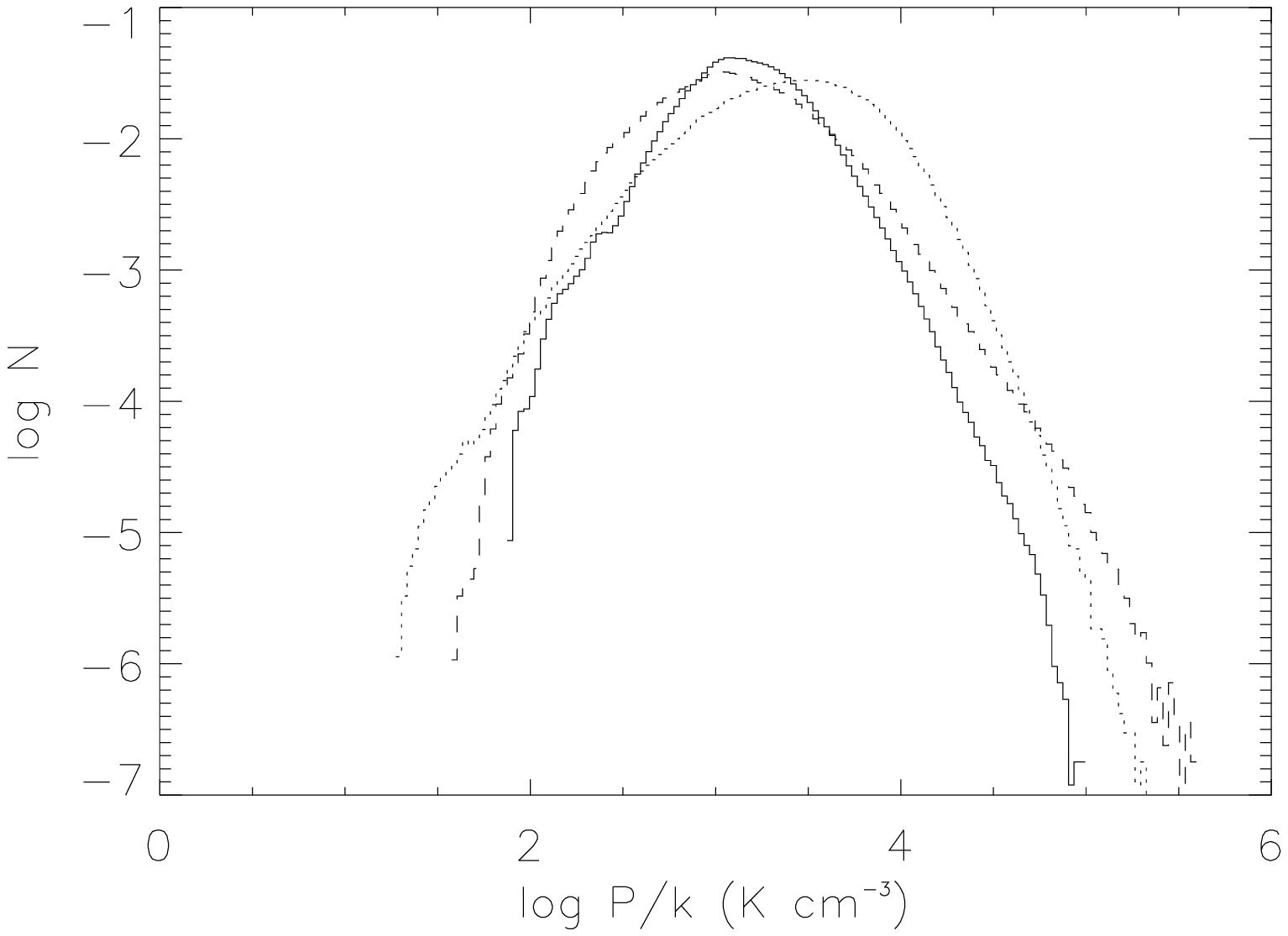}{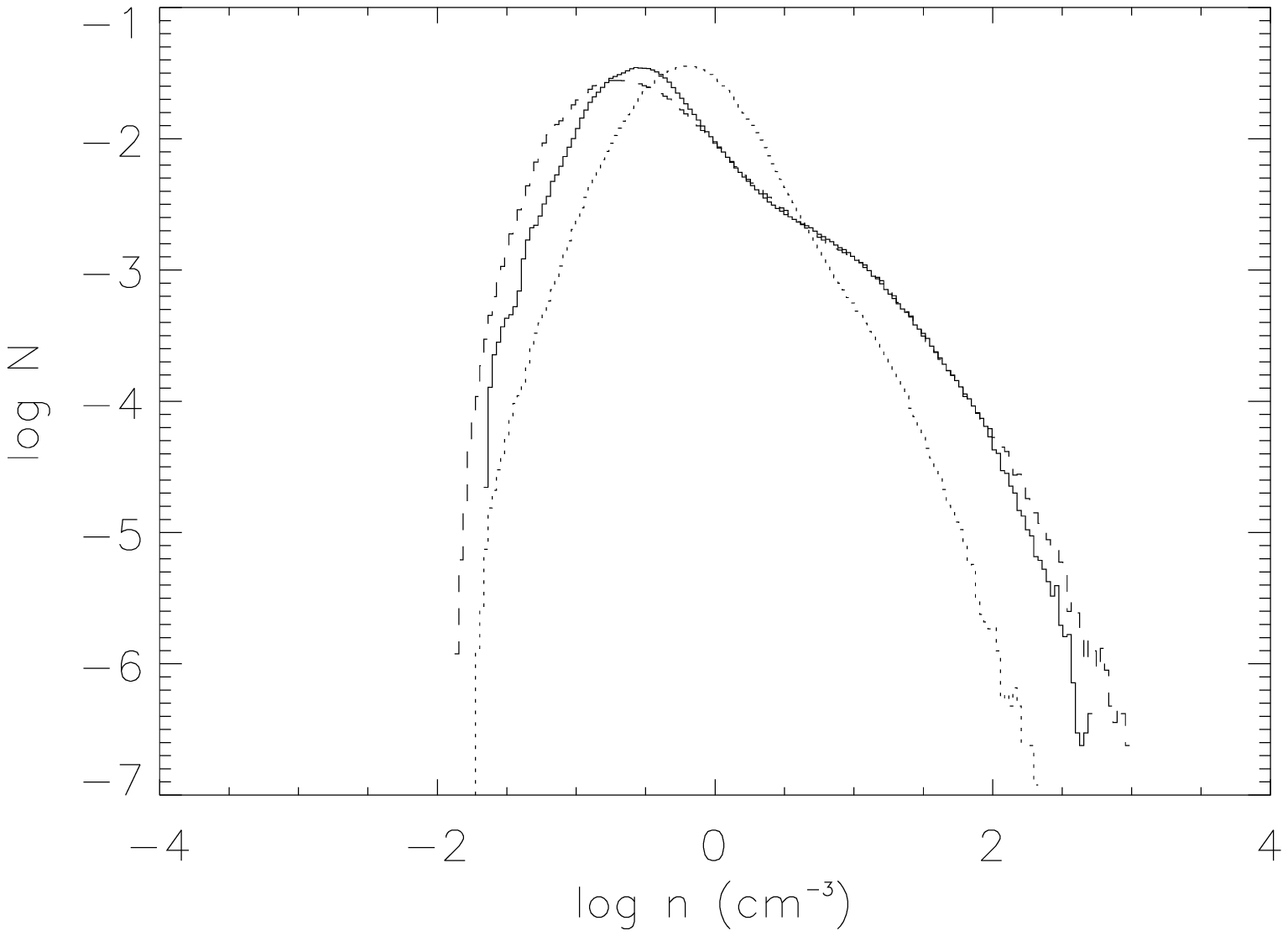}
\caption{Total pressure ({\it left}) and density ({\it right}) 
 histograms for 3D simulations with $M= 0.85$ and $\kfor=2$ ({\it solid line}), 
 $M= 0.85$ and $\kfor=8$ ({\it dotted line}), and $M= 1.35$ and $\kfor=2$ 
({\it dashed line}). The histograms are normalized to the
total number of points.} 
\label{fig:trends3d}
\end{figure}

The trends described in previous sections for the pressure 
and density distributions resulting from 2D simulations as $M$ and
$\kfor$ are varied are also maintained in the 3D case. 
In fig.\ \ref{fig:trends3d}a and b we respectively show the pressure and
density histograms resulting  
from three simulations with $(M,\kfor)=(0.85,2)$, $(0.90,8)$, and $(1.35,2)$.
It can be observed that as $\kfor$ increases, the width  of the pressure 
distribution
and  the value of $P(N_{\rm max})$ 
increase while the density histogram becomes 
narrower and with a less pronounced bimodality. On the other hand, when the value
of $M$ is increased, the pressure distribution widens noticeably, while the density 
distribution widens marginally; also, the bimodality of the latter
is slightly less pronounced. 

Finally, the fraction of cells with $n > 7.1$ cm$^{-3}$ and $P \geq 10^4$ K
cm$^{-3}$ for the 3D simulation with $\kfor=2$ and $M=0.85$ is
1.3\%, while that obtained in the 2D run with $\kfor=2$
and $M=0.97$ is 1.2\%, again showing a high consistency between the 2D
and 3D cases.

We conclude that the results obtained from the 2D simulations presented in
\S\S \ref{subsec:scale} and \ref{subsec:mach}, as well as the discussion from
\S \ref{sec:discussion}, still hold in 3D.

\subsection{Relation to previous work} \label{sec:implications}

Our results support the scenario that in a turbulent, thermally bistable
flow, there exists a fraction of fluid parcels that are out of thermal
equilibrium, and that this fraction depends on the local ratio $\eta$ of the
turbulent crossing time to the cooling time. Recently, Wolfire et
al.\ (2003) have estimated this ratio for the warm neutral medium
(WNM), using an approximation for the 
turbulent crossing time given by  $t_{\rm shock} \sim \ls/c_{\rm
s}$ (i.e., the ``mean time between shocks''),
where $c_{\rm s}$ is the sound speed and $\ls$ is the scale at which
the typical turbulent velocity 
difference equals $c_{\rm s}$. We refer to $\ls$ as the ``sonic''
scale. Wolfire et al.\ (2003) estimated $\ls \sim 200$ pc for the WNM,
and $\sim 0.3$ pc for the \emph{cold} neutral medium (CNM), although they 
warned that this is a very uncertain quantity. With these estimates,
they found 
$\eta \sim 0.3$--0.9 for the WNM, concluding that non-equilibrium
temperatures should often be found in the WNM, in agreement with
observations (Dickey et al.\ 1977;
Kalberla et al.\ 1985;  Spitzer \& Fitzpatrick 1995; Fitzpatrick \&
Spitzer 1997; Heiles 2001; Heiles \& Troland 2003; Kanekar et
al. 2003) and previous numerical studies (Gazol et al.\ 2001).

This conclusion is also consistent with our present results. Adopting
their value 
of $\sim 200$ pc for $\ls$, and a Kolmogorov velocity dispersion scaling
law $\Delta v \propto \lambda^{1/3}$, appropriate for incompressible
turbulence, we see that the rms Mach number with respect to the WNM at a
scale of our simulation boxes (100 pc), should be $\sim
2^{-1/3} \approx 0.8$. Thus, based on their estimates, the WNM on
scales of 100 pc should be bracketed by our simulations
with $M = 0.5$ and $M=1$, at the largest driving scales ($\kfor =2$),
with the $M=1$ case being the most relevant.

However, it is possible that the above regime, arrived at through the
considerations of Wolfire et al.\ (2003), still somewhat
underestimates the role of turbulence. Those authors noted that,
because of the scaling of the turbulent velocity with size, 
below the sonic scale the medium should roughly be in pressure
equilibrium. However, this does not guarantee that thermal instability
will be fully unimpeded in the generation of the density structures
below this scale. As discussed by
\VS\ et al.\ (2003), in a turbulent, thermally-bistable flow, there are
three competing timescales: the sound crossing time $\ts$, the turbulent
crossing time $\ttu$, and the cooling time $\tc$. In the standard linear
analysis (Field 1965), the largest growth rate of the instability is
given by the inverse of the cooling time, which is independent of scale
in the linear regime. This largest growth rate occurs at an
intermediate-wavelength regime $\lambda_{\rm F} \ll \lambda \lesssim
\Leq$, where $\lambda_{\rm F}$ is the Field length and $\Leq$ is the
scale at which $\ts \sim \tc$ (c.f.\ \S \ref{sec:discussion}). 
On the other hand, the sound and turbulent crossing times do
depend on scale, with $\ts \propto \lambda$ and $\ttu \propto
\lambda^{2/3}$ (assuming an incompressible Kolmogorov spectrum), or $\ttu
\propto \lambda^{1/2}$ (assuming a highly compressible, Burgers-like
spectrum, appropriate for the unstable range). So, even 
though indeed  the ratio of sound-to-turbulent 
crossing times becomes progressively smaller with decreasing scale size,
the ratio $\eta \equiv \ttu/\tc$ also becomes smaller. This suggests that
nearly incompressible, shearing turbulent fluctuations may have time to
disrupt TI-induced condensations before they grow, at least
partially. This effect may be enhanced in the presence of magnetic
fields, which increase the solenoidal (shearing) fraction of the
turbulent kinetic energy (\VS\ et al.\ 1996). Indeed, even in our
weakest-turbulence (largest-$\eta$) simulations ($M=0.5$, $\kfor=2$),
the bimodality of the density PDF (caused by the thermal bistability) is
moderate (fig.\ \ref{fig:histm2} \emph{right panel}), and the global
effective polytropic exponent is already positive across the thermally
unstable range (fig.\ \ref{fig:slopes_hiPfrac}, \emph{left panel})).

Our results also have the implication that the development of a high-$P$
power-law tail is not exclusive 
to supernova-driven models, such as the McKee \& Ostriker (1977) model,
and that pure
advective driving can generate approximate power-law tails in low-$\gamef$
flows at sufficiently high Mach numbers. 
On the other hand, noting that we also obtain lognormal-like 
shapes in the high-$P$ sides of the PDFs for large $\kfor$ suggests that the
formation (or not) of power-law tails in the 
pressure PDF depends on $\gamef$.

Finally, the fact that the pressure histogram partially copies features
of the density histogram, in particular the production of power-law
tails as a function of $\gamef$, gives validation to the approach used
by Mac Low et al.\ (2004), who applied a change of variable from density
to pressure in order to understand the functional form of the pressure
PDF from the known shape of the density PDF. On first account, such a
procedure might be questioned because of the large scatter exhibited by
the pressure at any given density, but our results suggest that it may
be valid on average.

Observationally, Jenkins (2004) has recently presented histograms of the
pressure in observations of CI, which can in principle be compared with
our results, as well as with pressure histograms from full ISM
simulations, like those of Mac Low et al.\ (2004) and de Avillez \&
Breitschwerdt (2004). However, at this point such a comparison is ambiguous,
because Jenkins (2004) had to \emph{assume} an effective polytropic
exponent in order to correct for an over-representation of dense points
in his observational sample. Since our results indicate that
$\gamef$ determines the shape of the pressure histogram, it appears
difficult to disentangle this physical role of $\gamef$ from its
possible role in biasing the observed pressure distribution. A detailed
attempt to perform such a comparison will be presented elsewhere, but
here we just note that our fraction of points with $P \geq 10^4$ K
cm$^{-3}$ for the case $M=0.5$, $\kfor=2$, of order 0.07\%, appears
consistent with the fraction reported by Jenkins (2004) for approximate
thermal equilibrium in the cold medium, also $\sim 0.1$\%.

\section{Conclusions}\label{sec:conc}

In this paper we have carried out a systematic study of the effect of
turbulent velocity fluctuations, characterized by their rms Mach number
$M$ and the energy injection wavenumber, $\kfor$, on the thermal
pressure distribution in a thermally bistable flow. To this end we have
performed a large number of 2D simulations varying those two parameters
in 100-pc boxes with random turbulent driving
generated in Fourier space, which allows precise control of the
parameters. A few test cases in 3D suggest that the 2D results are valid
in 3D as well. Our results are consistent with the picture that
as either of these parameters is increased, the ratio of turbulent
crossing time to cooling time decreases, causing a departure from
thermal equilibrium, and an approach towards an \emph{adiabatic}
behavior. This translates into an increase of the effective polytropic
index $\gamef$, as measured by the slope of the distribution of points
in the pressure-density diagram, in turn creating a population
of underpressured  zones in the 
diffuse gas, and of overpressured zones in the dense gas, with respect
to the thermal-equilibrium value of the pressure, $\Peq$. In particular,
the fraction of zones with densities $n > 7.1$ cm$^{-3}$ and with $P
> 10^4$ K cm$^{-3}$ increases 
from roughly 0.1\% at $\kfor =2$ (a driving scale of 50 pc) and $M=0.5$ to
roughly 70\% for $\kfor=16$ (a driving scale of 6.25 pc) and
$M=1.25$. In particular, for $M=1$ and $\kfor=2$, this fraction is
$\sim 1$\%, similar to the value reported by Jenkins (2004) from a
survey of the fine-structure excitation of CI on the Glactic plane.

In all cases, the dynamic range of the pressure in any given density
interval is larger than one order of magnitude, and the total dynamic
range, summed over the entire density range, typically spans 3--4 orders
of magnitude. The total pressure histogram widens as the Mach number is
increased, and moreover develops near-power-law tails at high (resp.\
low) pressures when $\gamef \lesssim 0.5$ (resp.\ $\gamef \gtrsim
1$), which occurs at $\kfor =2$ (resp.\ $\kfor=16$) in our
simulations. The opposite side of the pressure histogram decays rapidly,
in an approximately lognormal form. 
The development of power-law tails in the
pressure PDF is analogous to, and in fact more pronounced than, that
observed in the density PDF, suggesting that the
average value of the pressure is sufficiently correlated with the
density as to reflect the same dependence of its histogram with
$\gamef$, in spite of the large scatter of the pressure at any given
density. This may validate approaches attempting to obtain the pressure
PDF from that of the density via a change of variables.

\acknowledgements
We would like to acknowledge E. B. Jenkins and P. Hennebelle for useful
comments and thank an anonymous referee for suggestions
to improve this work.
This work has received partial financial support from CONACYT grant
36571-E to E.\ \VS\ and from DGAPA grant PAPIIT-IN114802 to A. Gazol. 
The work of J.\ Kim was supported by the Astrophysical Research Center for the
Structure and Evolution of the Cosmos (ARCSEC) of Korea Science and Engineering
Foundation (KOSEF) through the Science Research Center (SRC) program.
The numerical simulations were performed on the
linux clusters at CRyA (funded by CONACYT grant 36571-E) and at
KAO (funded by KAO and ARCSEC). This work has made extensive use of
NASA's Astrophysics Abstract Data Service and LANL's astro-ph archives.

\clearpage

\clearpage

\clearpage

\clearpage

\clearpage

\epsscale{1.0}

\clearpage

\clearpage

\epsscale{1.0}

\clearpage

\epsscale{1.0}

\clearpage

\epsscale{1.0}

\clearpage

\epsscale{1.0}

\clearpage

\clearpage

\clearpage

\clearpage


\clearpage

\epsscale{1.0}

\clearpage

\clearpage


\end{document}